\documentclass[fleqn,usenatbib]{mnras}

\usepackage{newtxtext,newtxmath}

\usepackage[T1]{fontenc}
\usepackage{ae,aecompl}

\usepackage{graphics,graphicx}	
\usepackage{amsmath}	
\usepackage{amssymb}	
\usepackage{lineno}
\usepackage{float}
\DeclareGraphicsExtensions{.png,.pdf,.jpg}

\newcommand{\xte}{XTE~J1550$-$564}
\newcommand{\Cha}{{\it Chandra~}}

\newcommand{\ergcm}{erg cm$^{-2}$ s$^{-1}$}

\newcommand{\dg}{$^\circ$}

\newcommand{\er}{$\pm$}

\title[The \xte~western jet in radio and X-rays]{Evolving Morphology of the Large-Scale Relativistic Jets from XTE J1550$-$564}

\author[G. Migliori et al.]{
Giulia Migliori$^{1}$\thanks{E-mail: giulia.migliori@cea.fr},
S. Corbel$^{1,2}$,
J.~A.~Tomsick$^{3}$,
P.~Kaaret$^{4}$,
R.~P.~Fender$^{5}$,
\newauthor
A.~K.~Tzioumis$^{6}$,
M.~Coriat$^{7,8}$,
J.~A.~Orosz$^{9}$
\\
$^{1}$Laboratoire AIM (CEA/IRFU - CNRS/INSU - Universit\'e Paris Diderot), CEA DSM/IRFU/DAp, F-91191 Gif-sur-Yvette, France.\\
$^{2}$ Station de Radioastronomie de Nançay, Observatoire de Paris, PSL Research University, CNRS, Univ. Orléans, 18330 Nançay, France.\\
$^{3}$Space Sciences Laboratory, 7 Gauss Way, University of California, Berkeley, CA 94720-7450, USA.\\
$^{4}$Department of Physics and Astronomy, University of Iowa, Iowa City, IA 52242, USA.\\
$^{5}$Astrophysics, Department of Physics, University of Oxford, Keble Road, Oxford OX1 3RH, United Kingdom.\\
$^{6}$Australia Telescope National Facility, CSIRO, P.O. Box 76, Epping NSW 1710, Australia.\\
$^{7}$Universit\'e de Toulouse, UPS-OMP, IRAP, Toulouse, France.\\
$^{8}$CNRS, IRAP, 9 av. Colonel Roche, BP 44346, F-31028 Toulouse cedex 4, France.\\
$^{9}$Department of Astronomy, San Diego State University, 5500 Campanile Drive, San Diego, CA 92182-1221, USA.}
\date{Accepted XXX. Received YYY; in original form ZZZ}

\pubyear{2017}

\begin{document}
\label{firstpage}
\pagerange{\pageref{firstpage}--\pageref{lastpage}}
\maketitle

\begin{abstract}
We present an in-depth study of the large-scale, western jet of the microquasar \xte, based on X-ray and radio observations performed in 2002-2003. The jet is spatially resolved in both observing windows. The X-ray jet is expanding in time along the axis of the jet's propagation: we observe the formation of a tail ($\sim$2.25\arcsec), which appears to extend backwards with an apparent velocity $\sim-$0.10$c$. The origin of this feature is discussed in the framework of scenarios of energy dissipation. A single power-law adequately describes the broadband spectra, supporting a synchrotron origin of the X-ray emission. 
However, a spectral break at $\approx$10$^{15}$ Hz is necessary in coincidence with a re-flare at 8.64 GHz in September 2002. This finding may be indicative of emission from newly accelerated low-energy particles. The first detection of the jet is in February 2001 (F$_{\rm 8.64GHz}=$0.25\er0.09 mJy) in the flux rising phase. A phase of stable emission is followed by a rapid decay ($t_{\rm decay}=$167\er5 days). The decay at radio frequencies is significantly shorter than in X-rays ($t_{\rm decay}=$338\er14 days). We detected a high fraction (up to $\sim$9\%) of linearly polarized radiation at 4.8 GHz and 8.6 GHz. The orientation of the electric vector is consistent with the picture of a shock-compressed magnetic field, and there are hints of variations on month-timescales, possibly connected with the evolution of the jet structure.

\end{abstract}




\section{Introduction}
\label{intro}
In Galactic stellar-mass X-ray binaries (XRBs) jets are observed in different flavors: compact jets and relativistic jets at the small scales, diffuse and persistent shells and lobes or transient hot spots at the large scales. Each flavor contributes to probe important aspects of the physics of jets, such as their formation, their structure and energetics, and their impact on the interstellar medium (ISM).

At the very small ($\sim$10$^{-4}$ pc) scales,  compact jets produce flat, or inverted, synchrotron radio spectra \citep[][]{Cor00,Fen01,Fen06}. The shape of the radio spectrum points to self-absorbed synchrotron emission and it requires, in order to be maintained, a mechanism able to compensate for the energy losses due to the jet's adiabatic expansion \citep[][]{BK79}.
A possibility is that internal shocks mediate the conversion of a fraction of the kinetic energy of the jet into internal energy, which is then radiatively dissipated.
The shocks could arise from the collision of discrete shells of plasma traveling at different velocities \citep{KSS00,JFK10,Mal14}, in analogy with models proposed for the prompt emission of gamma-ray bursts (GRBs) and active galactic nuclei \citep[AGNs, see ][]{Spa01}. 

Transient bright jets with apparent superluminal velocities were discovered in XRBs on scales of 0.02-0.06 pc \citep[GRS 1915+105, GRO J1655$-$40][]{MR94,Tin95,Hje95}. 
The emission is coming from optically thin plasma moving with inferred actual velocities greater than $>$0.9$c$.  Because of similarities with AGNs, the definition of microquasars was coined for this class of XRBs.

At large scales, microquasars display both diffuse, persistent emission and transient jets or hot spots. The former is produced by structures which may extend over tens of parsecs and have been formed by recurrent radio outbursts \citep[e.g. 1E~1740.7$-$2942, GRS~1758$-$258][]{MR99}. In Cygnus X-1, the ISM gas has been shock-heated by the expansion of the radio lobe, which remains invisible, and formed a shell of thermal emission visible in the radio and optical bands \citep[][]{Gal05,Rus07}. 
X-ray emission marking jet-driven shocks have been reported in Circinus X-1 \citep{Sel10}, while in SS433 the jet is likely deforming and inflating the radio nebula W50 \citep{Dub98,Mig02}. Similarly to the giant lobes of radio galaxies, these structures store the power deposited by the jets over several radio outbursts, hence they are crucial to investigate the long term activity of XRBs/microquasars, and to estimate the average power of the jet, as well as the fraction of mechanical energy released into the ISM.

The detection of the transient, large scale (0.2-0.7 pc) jets is challenging. They may travel unseen for years after the outburst that has produced them, before being revealed through the interaction with the ambient medium. Furthermore, their emission fades quickly. For these reasons, so far only a handful of large scale jets have been found \citep[\xte, GX 339$-$4, H 1743$-$322, 4U~1755$-$33 and possibly XTE J1752$-$223][]{Cor02,Gal04,Cor05,Kaa06,Rat12}. Nonetheless, also by virtue of their rapid evolution, these transient jets are unique laboratories to unveil the structure and dynamics of the jets of microquasars, and of their supermassive counterparts, i.e. jetted active galactic nuclei (AGN).  The measured proper motions of  \xte~and H 1743$-$322 indicate strong deceleration of the ejecta with respect to the initial superluminal apparent velocities. In addition, monitoring observations show a gradual, yet continuous deceleration of the jets on month timescales. They produce decaying radio to X-ray broadband spectra. The presence of X-ray emission, which is most likely of synchrotron origin, points to a mechanism able to efficiently accelerate particles to $\sim$TeV energies at relatively large distances from the central source \citep{Tom03,Cor05,Kaa06}. 

\subsection{XTE J1550$-$564}
\xte~is a low-mass XRB at a distance of $\sim$4.4 kpc \citep{Oro11}, including a black hole (BH) primary with an estimated mass 9.1\er0.6 M$_{\sun}$ \citep{Oro11}. The orbital period is 1.54 days and the inclination of the orbital plane is $i=$75\dg\er4\dg~\citep{Oro11}. 
A pair of large scale ($\sim$0.6 pc) transient jets were identified in the radio and X-ray bands \citep{Cor02,Tom03,Kaa03}. Their formation was related to a major X-ray outburst of the central source, which reached the peak of the emission in September 19-20, 1998 \citep{Sob00}, when the source was likely at its Eddington limit \citep{Ste11}. The X-ray outburst was accompanied by the ejection of bipolar relativistic jets, observed in two epochs by the Australian Long Baseline Array (LBA) at an angular separation of 255$^{+15}_{-19}$ mas \citep{Han01,Han09}, which,  at the distance of \xte, implies an apparent separation velocity and an intrinsic velocity of 1.7$c$ and $\gtrsim$0.9$c$, respectively \citep{Oro11}.

At larger scales, the eastern jet was detected first, in April 2000, indicating that it is the approaching jet, at $\sim$20\arcsec~from the central source, while the receding western jet appeared in March 2002 and remained visible until October 2003. Monitoring with the \Cha X-ray telescope, between 2000 and 2003, allowed to trace the advancing motion of the large-scale X-ray jets and revealed a significant decrease of their apparent final velocities \citep[$\sim$1.0$c$ and $\sim$0.6$c$ for the eastern and western jets, respectively][]{Cor02}  with respect to the ejection epoch in 1998.  

The gradual, constant deceleration measured in the subsequent observations \citep{Cor02,Tom03}, together with the changing morphology of the western X-ray jet \citep{Kaa03}, point to a dynamic, evolving system. 
The X-ray emission of both jets is consistent with the extrapolation at high energies of their respective radio fluxes, supporting its synchrotron origin \citep{Cor02,Tom03,Kaa03}. 

The X-ray observations  were used to model the dynamics and emission of the large-scale jets \citep{Wan03,HZ09,SM12}. In the scenario emerging from the modeling, the jets propagated basically unhindered  through a cavity of low ISM density until they reached the cavity's walls. The jump in the ISM density profile triggered the formation of external shocks responsible for the acceleration of the particles which produced the radio-to-X-ray emission \citep{Wan03,HZ09}. 
 
The previous studies investigated the kinematics of the X-ray jets (Corbel et al. 2002, Tomsick et al. 2003, Kaaret et al. 2003, Hao \& Zhang 2009, Steiner et al. 2012). 
In this work, we present for the first time the results of the multi-frequency radio monitoring performed with the Australia Telescope Compact Array (ATCA). 
The primary target of the observations was the western jet, which remained visible in the radio band for a longer time than the eastern jet and displayed a spatially resolved, evolving morphology in X-rays \citep{Kaa03}. 
The ATCA and \Cha observations, which overlapped during a $\sim$1.5-year period, provide us with an unique multi-band time-lapse sequence of the evolving large scale western jet. We employed sub-pixel imaging of the \Cha~data in order to obtain X-ray images of the western jet at unprecedentedly high resolution. 
The ATCA observations show that the western radio jet first appeared in 2001, giving us the possibility to trace the jet's radio flux from its rise to its decay. We also report for the first time the detection of linearly polarized radio emission from the western large scale jet.

We present the ATCA observations in  Section~\ref{radio} and the X-ray analysis in Section~\ref{Xray}. The results on the morphology, spectra, polarization and variability of the western jet are reported in  Section~\ref{results}. We discuss the jet structure and the nature and origin of the observed emission in  Section~\ref{discussion} and summarize the main results of the work in  Section~\ref{summ}.

\section{Radio Observations}
\label{radio}
The ATCA synthesis telescope is located in Narrabri, New South Wales, Australia, and consists of an east-west array with six 22 m antennas. 
\xte~was observed by the ATCA with regularity from January 2002 until July 2003. Prior to this, seven more observations were sparsely performed between November 1999 and February 2001. In total, between November 11, 1999  and July 25, 2003 we have 31 radio observations of \xte with the ATCA, mostly at 4800 and 8640 MHz with different array configurations (see details in Table \ref{t1}).  Due to the rapid decay of the radio flux in the last two observations (obs23 and obs24), both observing frequency windows of ATCA were centered at 4.8 GHz in order to gain in flux sensitivity. Occasionally, observations at 1384 and 2496 MHz were also taken and allowed for a better characterization of the radio spectrum. A selection of these radio observations were presented in \citet{Cor02} and \cite{Kaa03}. The complete set of observations that is presented here (Table \ref{t1}) provided us with: 1- the radio light curve and spectra of the western jet over the full campaign period; 2- the radio morphology of the western jet from the observations with the best spatial resolution;  3- the radio positions of the western and eastern jets; and 4- measurements of the linear polarization of the western jet  at 4800 and 8640 MHz.

The amplitude and band-pass calibrator was PKS1934$-$638, while the antenna's gain and the polarization leakage were derived from observations of different nearby calibrators, B1554$-$64 was used at 4800 and 8640 MHz, while B1549$-$790 was used at 1384 and 2496 MHz. 
The editing, calibration, Fourier transformation, deconvolution, and image analysis were performed using the Multichannel Image Reconstruction, Image Analysis and Display (MIRIAD) software package \citep[][]{Sau95}.
 A natural weighting scheme was applied to the visibilities.
 
The western jet is detected from February 2001 until  July 2003, in either one of the four observing frequencies. The detection of the east jet is limited to the observations between April 2000 and February 2001. 

\section{X-ray Observations and Analysis}
\label{Xray} 
The western jet has been observed by {\it Chandra} with the Advanced CCD Imaging Spectrometer (ACIS) detector five times between March 2002 and October 2003. In all cases the source was placed on the back-illuminated ACIS S3 chip, which provides the best low-energy response. Before that, three observations performed in 2000 had led to the X-ray detection of the eastern jet \citep{Cor02,Tom03,Kaa03,HZ09,SM12}.
Given that the main goal of the X-ray part of this study is an in-depth analysis of the resolved structure of the western jet, here we re-analyzed the five 2002-2003 archival \Cha observations (see Table \ref{t2}).

The X-ray data analysis was performed with the Chandra Interactive Analysis of Observation (CIAO) 4.8 software \citep{Fru06} using the calibration files CALDB version 4.7.2. We ran the \texttt{chandra$\_$repro} reprocessing script, that performs all the standard analysis steps. In order to fully exploit \Cha imaging capabilities, the data were reprocessed with the Energy Dependent Sub-pixel Event-Repositioning \citep[EDSER,][]{Li04} algorithm, which replaces the prior pixel randomization and allows for an investigation of small scale structures on sub-pixel scales. We verified by applying the  \texttt{make$\_$psf$\_$asymmetry$\_$region} tool\footnote{http://cxc.cfa.harvard.edu/ciao/caveats/psf$\_$artifact.html} that the asymmetry, which has been discovered in the ACIS point spread function (PSF) within the central arcsecond, does not affect our analysis.  We filtered the data for the time intervals of background flares: only minimal cuts were applied in all observations except for ObsID 4368 and ObsID 5190, during which high background flares were detected and excluded from the analysis. For each data set, we generated smoothed images in the 0.3-8.0 keV, 0.3-2.5 keV and 2.5-8.0 keV energy bands with half pixel resolution (0.246\arcsec) in order to study the morphology of the western jet (Figures \ref{f1} and \ref{f2}).

We used the CIAO \texttt{wavdetect} task to identify the X-ray sources in each 0.3-8 keV image, adopting a sequence of wavelet scales (i.e. 1, 2, 4, 8, 16 pixels) and a false-positive probability threshold of $10^{-6}$.  In all five observation, an X-ray source is detected at a position consistent with the western radio jet with a $\gtrsim$18$\sigma$ significance \citep[see also][]{Cor02,Kaa03,HZ09,SM12}. 
We tested the dependence of the \texttt{wavdetect} results to some of the parameter settings by varying e.g. the bin size (0.5 and 1, corresponding to the half and full pixel resolution, respectively) and the energy encircled fraction.
The centroid positions are not significantly affected by these parameters: the largest variation of the centroid position (0.47\arcsec)  was found in the last observation where the X-ray jet reaches its maximum extension. Alternatively, following \citet{Tom03}, the precision of the \texttt{wavdetect} centroid of the western jet was checked by measuring the centroid of a 16$\times$16 pixel (7.9\arcsec $\times$7.9\arcsec) square centered on the \texttt{wavdetect} position itself. The dimensions correspond to the maximum scale adopted in the \texttt{wavdetect} task. The largest offset, $\sim$0.8\arcsec~was found in the last two observations (ObsID 4368 and 5190). This is not worrying as this method is more appropriate for a compact source, whereas, in the last two observations, the jet emission extends beyond the square boundaries. Based on the results of the two methods, we assumed 0.5\arcsec~as the error on the centroid position.

For all five observations, we extracted an energy spectrum for  the western jet emission using the \texttt{specextract} routine, which produces the relative calibration response files. 
 The dimensions and eccentricity of the elliptical extraction regions were adapted to the changing shape and extension of the western jet.
 The background spectrum was extracted from an annulus centered on the western region with an inner radius of 8\arcsec~and an outer radius of 18\arcsec.
The X-ray spectral analysis was performed within Sherpa version 1 for CIAO 4.8 \citep[][]{Free01}. We fitted simultaneously source and background spectra. Given their limited number, the counts were not grouped and the spectra were fitted using the Cash statistics with the simplex-neldermead method. We used the results of the fits to run simulations using pyBLoCXS\footnote{pyBLoCXs is Markov chain Monte Carlo based algorithm to perform Bayesian low-count X-ray spectral analysis in Sherpa, see http://cxc.cfa.harvard.edu/sherpa4.9/threads/pyblocxs/.} and determine the uncertainties on the model parameters and fluxes.
Finally, we extracted separate spectra for the main peak and tail regions of the western jet. We defined the PSF of a point source at the peak location using the Chandra  Ray Tracer application \citep[{\it ChaRT},][]{Car03}. A simple power law was adopted as input model for the simulations. In this way we determined that $\gtrsim$90\% of the main peak emission is enclosed in a circular region of radius $\lesssim$1.6\arcsec. The region of the  tail spectrum was obtained by subtracting the peak region from the region of the total jet.
 Throughout the paper, uncertainties on the spectral parameters and on the fluxes are at 1$\sigma$ confidence level.

\section{Results}
\label{results}
\subsection{Western Jet Morphology}
\begin{figure*}
\centering
\includegraphics[scale=0.5]{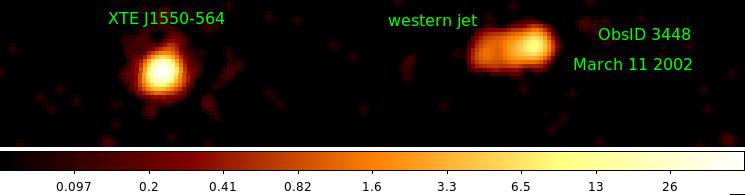}
\includegraphics[scale=0.5]{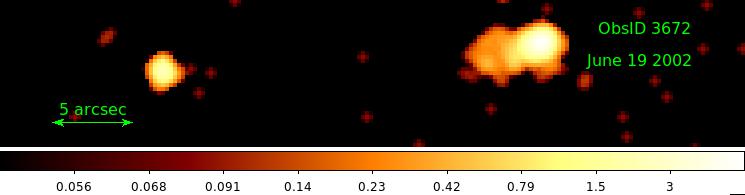}
\includegraphics[scale=0.5]{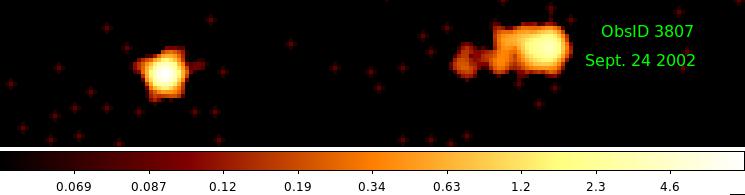}
\includegraphics[scale=0.5]{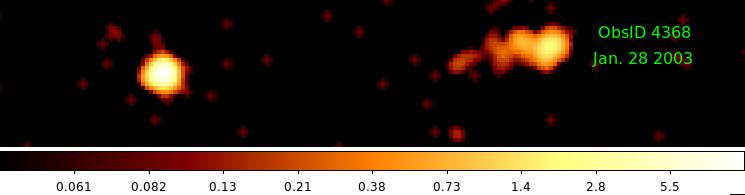}
\includegraphics[scale=0.5]{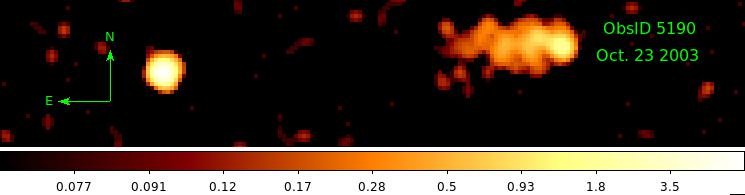}
\caption{Smoothed \Cha~0.3-8 keV ACIS-S images of the western jet of XTE J1550$-$564: \xte~is the source to the left and the western jet is located to the right. The pixel size is set to half of the original ACIS pixel and is equal to 0.246\arcsec. The five images are matched by R.A. and DEC and astrometric correction has been applied. The color scale, in count units, is logarithmic and the minimum of the scale is set to 3$\times$rms as measured in each image. For a direct comparison of the jet in the five observations see Figure \ref{f9}. }
\label{f1}
\end{figure*}

Since its first detections in March and June 2002, the X-ray emission of the western jet appeared to be extended in the direction of the main jet axis \citep{Kaa03}.
A 5\arcsec~extension characterized by a leading peak and a trailing tail was clearly visible in the X-ray brightness profile. 
The five 0.3-8.0 keV images in Figure \ref{f1} reveal that  X-ray morphology of the jet  is continuously evolving over the $\sim$1.5 years covered by the observations.
As the jet's brightness decreases, the emission becomes less uniform and compact while the brightness contrast between the leading peak and the trailing tail decreases. The final structure appears to be consistent with an helical shape. In Figure \ref{f2}, we investigated the distribution of the soft (0.3-2.5 keV) and hard (2.5-8 keV) X-ray emission within the jet. In the first observation, the morphology of the jet in the two bands is basically the same. In the last observation, the hard X-ray emission appears to be more knotty than the soft emission, but it is clearly present along the whole jet extension.  We performed simulations using MARX \citep{Dav12} to verify whether in the last \Cha observation the emission in the hard X-rays is truly more clustered than in the soft X-rays, once we account for the difference of counts (88 counts and 46 counts in the soft and hard band, respectively). The simulations (see the Appendix for the details) show a more uniform spatial distribution of the 2.5-8 keV emission for a number of counts comparable to that in the soft band (in Figure \ref{f2app}). Conversely, the 0.3-2.5 keV emission appears more knotty when the counts are reduced to $\sim$50. Based on these simulations, we cannot argue for a different spatial distribution of the emission in the two bandpasses. Common to the simulated jets in the two bands is instead the tendency of the photons to preferentially accumulate along the central spine, rather than on the external  borders. 

\subsubsection{X-ray brightness profiles}
 In order to further study the morphology of the western jet, we analyzed its X-ray emission. To evaluate the jet extension, we followed the method applied by \citet{Kaa03} in the first two \Cha~detections.  
We considered the count distribution in the 0.3-8 keV band along the axis connecting the centroid of XTE J1550$-$564 to that of the western jet. The width of the box region used to collect the counts was slightly adjusted each time in order to include the whole jet and optimize the signal to noise ratio. For each observation, we obtained longitudinal profiles with bin sizes of 0.25\arcsec (Figure \ref{f4}). 
We measured the background counts per bin in regions which are located close to our source and free from point sources.
Given that the maximum background ($<$0.5 cts/bin in ObsID 5190 for a bin width of 0.25\arcsec) was always very low, we neglected it for the rest of the analysis.

First, we assessed the significance of the jet extension. We compared the profile of the western jet with the profile of XTE J1550$-$564, which is a point-like source and can be used as a calibration of the point spread function\footnote{This approach is alternative but consistent with the PSF simulations produced with {\it ChaRT} (Sec. 3). Note that the change in shape and size of the PSF due to the different locations in the telescope field-of-view and the spectral energy distributions  of \xte~and of the western jet should be negligible: http://cxc.harvard.edu/ciao/PSFs/psf$\_$central.html. Furthermore, the instrument settings were defined so that the PSF was not affected by pile-up effects even when \xte~was the brightest \citep{Kaa03}. }.  The XTE J1550$-$564 profile was rescaled and moved to match the position of the peak of the western jet. The point source profile reproduces the front edge of the emission, however, confirming what is seen in the images, in all the five profiles there is evidence of an excess of counts forming a tail in the direction of the central source. A Kolmogorov-Smirnov (K-S) test between the profiles of the western jet and of the \xte~confirms and strengthens the results for the first two observations of \citet{Kaa03}.  In the last three observations, the null probability, i.e. the probability that the two samples are drawn from the same distribution, is $<$5$\times 10^{-5}$, with a maximum difference between the cumulative probability distributions D$>0.58$.
These results hold even when we increase the bin size of the brightness profile to 0.5\arcsec. 

Next, we investigated the evolution in time of the western jet profile.  A K-S test indicates that the profiles of the western jet during its first (March 2002, ObsID 3448) and last (October 2003, ObsID 5190) detections are inconsistent at the 99\% confidence level.  In order to follow the evolution of the jet, we measured the angular separation of its main peak and of the trailing edge from  XTE J1550$-$564 (i.e. their respective projected distances from \xte~along the jet axis).
We defined the  end point of the tail as the last bin with at least two counts. In addition, given the low background level, we also reported the position of the last one-count bin that is part of a group of at least 5 photons (within 1.5\arcsec) and, by visual inspection, appears related to the western jet. 
We applied the astrometric correction to the data, so that in all \Cha~observations the X-ray centroid of the central source coincides with its radio position \citep[][]{Cor01}. 
The orientation of the jet axis in the five observations varies by less than 0.1\dg. This means that the advancing direction of the jet does not change significantly in time and makes possible to compare the projected positions of the jet features among the five observations. 
The positions of the peak and of the tail are reported in Table \ref{t2}.
For a bin size of 0.25\arcsec, the main peak distance from the central source increases from 22.75\arcsec~in the first observation to 24.5\arcsec~in the last one. We note that the peak of the emission of the western jet and its centroid, although not necessarily coincident, are consistent within the uncertainties.
In the same period the distance between the tail edge and \xte~appears to decrease by 2.25\arcsec. Thus, we find that while the front of the jet is advancing, its rear part appears to move backwards.

We investigate the possibility that the backward expansion of the tail is an apparent effect of the different exposures of the observations.
To this purpose, for each observation we measured the X-ray flux in a circular region of 1.4\arcsec~radius. The region position was selected so that it contained the tail emission as observed in ObsID 5190 (RA=15h 50m 56s.57, DEC=$-$56$^\circ$ 28$^\prime$ 33.55\arcsec), while it appeared free from the jet emission in the first observation. 
 We used the \texttt{srcflux} script to obtain the 0.5-8.0 keV count rates and fluxes in the selected region in all the observations\footnote{Note that the astrometric correction was applied to the data.}. Input parameters of the script were the extraction region of the source and of the background. An absorbed power law was assumed as a spectral model. The photon index was fixed to the best fit value obtained from the spectral analysis (see Sec. 4.3) and the absorbing column was fixed to the Galactic value. The bin size value was set to 0.5 (corresponding to the sub pixel size, $\sim$0.25\arcsec). The psf parameter was set to ideal (as suggested in the case of extended emission). 
X-ray emission in the selected region is detected with a $>$3$\sigma$ significance only in the last observation, with a count rate of (1.6$^{+1.2}_{-0.8}$)$\times 10^{-4}$ c/s and a flux of (2.4$^{+1.8}_{-1.2}$)$\times 10^{-15}$ \ergcm. Upper limits of 2.1$\times 10^{-4}$ c/s are measured in the two observations of September 2002 and June 2003 (ObsID 3807 and ObsID 4362 respectively). This value is still compatible with the tail signal measured in the last observation.
In the first two observations in March and June 2002 (ObsID 3448 and 3672), when  the X-ray jet was the brightest, the count rate upper limits are  1.5$\times 10^{-4}$ c/s and 1.3$\times 10^{-4}$ c/s, respectively, i.e. below the count rate measured in ObsID 5190. The non-detection of the X-ray tail in the first two observations gives support to the backward expansion of the tail. 
 
To test and quantify the significance of the expanding X-ray tail, we performed simulations of the X-ray jet with MARX and Sherpa (see the Appendix for details). The simulations confirmed that the jet length significantly increased from the first to the last \Cha~observation. We adopted the standard deviation on the jet's length of the simulated jets (0.4\arcsec, see Table \ref{t6}) as the error on the tail's endpoint. By adding in quadrature the errors, we obtained a statistical uncertainty of 0.6\arcsec~on the measurements of the tail expansion between the first and last detection.

 Our simulations showed that the extension of tail is sensitive to the different exposure time of the observations (see the discussion in the Appendix). On one hand, we concluded that the exposure time of the first \Cha~observation was sufficient to constrain the morphology of the western jet, given its brightness and compactness.  On the other hand, the actual extension of the extended jet in the last detection is more uncertain: simulations of deep (100 ksec) \Cha~observations suggest that we might be underestimating its real extension.  While it is difficult to quantify this uncertainty, the reader should bear it in mind in the discussion relative to the tail motion (Sec. 5.1).   

The distribution of the X-ray counts of the western jet, in the direction perpendicular to the main axis, was found to possibly be extended ($<$0.8\arcsec) in the first two \Cha~observations \citep{Kaa03}. We considered the remaining three observations. The vertical profiles included the 0.3-8 keV counts with \er2\arcsec displacement from the jet's main axis in the region between  the jet peak and the tail positions (see in Table \ref{t2}). A K-S test shows that in each observation the perpendicular profiles of the western jet and \xte~are not drawn from the same parent population at the 99\% confidence level (see Figure \ref{f4a} for a comparison of the two profiles in the last \Cha~observation). We find indications of a change of the perpendicular profiles between the first and the last \Cha~observations (97\% confidence level, see Figure \ref{f4a}). The standard deviations of the distributions of the X-ray counts in ObsID 3448 and 5190 are 0.48\arcsec\er0.01 and 0.79\arcsec\er0.02, respectively. We repeated the analysis slicing the jet in two regions, so to separately compare the leading edge (the counts at \er2\arcsec from the centroid position along the jet's displacement axis)  and the tail component in the two observations: while the width of the former is basically unchanged in time, the tail appears to be more spread in the last observation (96\% confidence level), with the standard deviations of the distributions of the X-ray counts being 0.66\arcsec\er0.02 and 1.00\arcsec\er0.04 in ObsID 3448 and 5190, respectively.

\subsubsection{Radio morphology}
The ATCA observations at 4.8 and 8.6 GHz allow for a comparison between the X-ray and radio morphologies of the western jet.  
 We selected the radio observations which were taken with the most extended array configuration (6 km array, see Table \ref{t1}) and the closest in time to the X-ray observations. We used radio maps at 8.6 GHz frequency, which have the best spatial resolution, except for the  the last X-ray observation of October 2003, when only a 4.8 GHz image was available. 
 In Figure \ref{f9}, the radio contours are overlapped to the X-ray images.
Despite the resolution of the radio images is not as good as in X-rays, we observed a change of the jet morphology also in this band.
The position of the main radio peak is consistent with that of the X-ray peak in the first four \Cha~observations, although a slight shift between the two peaks is evident in September 2002 (ObsID 3807 and obs17  in X-ray and radio, respectively). The radio emission is also spatially extended, although to a less degree than in X-rays.
In the last X-ray  observation, the morphologies in the two bands appear rather different and the peak of the radio emission is shifted downstream with respect to the X-rays.  However, in this last case the time interval between the radio and X-ray observations is the largest ($\sim$3 months).  Assuming the apparent advance velocity of the jet that we measured in observation 24 ($v_{\rm app.,obs1/01}=$3.1\er1.0 mas day$^{-1}$, see Sec. \ref{propmot} and Table \ref{t4radsep}), by the time of the last X-ray observation the radio peak could have shifted of $\lesssim$0.4\arcsec , becoming consistent with the position of the X-ray peak.

In addition to these selected observations, the emission of the radio jet is extended in the majority of the detections at 8.6 and/or 4.8 GHz that were obtained with the extended array configuration. A fit of the 8.6 GHz jet flux with a single elliptical Gaussian left significant residuals and in more than half of the observations we identified at least one secondary component (see Table \ref{t1}).

The radio structure is not simply reproducing the X-ray morphology. In September 2002 the radio and X-ray morphologies are markedly different, with the former surrounding the northern border of the latter. In January 2003, the radio jet has an arch-like shape, partially resembling the complex structure visible in X-rays. 

\subsection{Eastern Jet Detection}
The eastern X-ray jet was clearly detected in four epochs \citep[June 2000 to March 2002, \Cha observations ObsID 679, 1845, 1846, 3448,][]{Cor02,Tom03,Kaa03}. A marginal detection is reported in September 2002 (ObsID 3807) by \citet{HZ09} and \citet{SM12}. In the X-ray images, emission compatible with the location of the eastern jet is indeed visible in ObsID 3807 ($\sim$17 counts), but also in the last \Cha observation in October 2003, (ObsID 5190, $\sim$13 counts). We used the CIAO script \texttt{srcflux} to evaluate the significance and the intensity of this signal. Given the low number of counts we set the bin size parameter equal to 1 (1 pixel=0.492\arcsec). In ObsID 3807 the eastern jet is detected with a significance 3.6$\sigma$ and a 0.5-8.0 keV flux of 8$^{+4}_{-3}\times 10^{-15}$ \ergcm~(equivalent to a count rate of 6$^{+3}_{-2}\times 10^{-4}$ c/s). In ObsID 5190 we obtained a marginal detection (2.95$\sigma$) and measured a flux of 3.4$^{+2.3}_{-1.6}\times10^{-15}$ \ergcm~(equivalent to a count rate of 2.3$^{+1.5}_{-1.1}\times10^{-4}$ c/s). The distance of the eastern jet's centroid (measured using the ds9 Funtools) from the central source is 28.6\arcsec\er1.0\arcsec~and 33.5\arcsec\er1.4\arcsec~in ObsID 3807 and 5190, respectively. Note that the distance measured in ObsID 3807  is smaller than the values of 29.2\arcsec\er1.4\arcsec~and 29.6\arcsec\er0.6\arcsec~in \citet{HZ09} and \citet{SM12}, respectively, although still in agreement within the (large) uncertainties.

\subsection{X-ray Spectra}
For all five observations, an absorbed power-law model reproduces the spectrum of the western jet (see Table \ref{t3spec}). 
If left free to vary, the absorption column density is in the range of $N_{\rm H}=(0.6 -1.3)\times10^{22}$ cm$^{-2}$, with the best fit values clustering around the Galactic value,  $N_{\rm H,Gal}=0.9\times10^{22}$ cm$^{-2}$. Therefore, although the presence of an intrinsic absorber is not excluded, for the rest of the analysis we fixed the $N_{H}$ to the $N_{\rm H,Gal}$.
The results of the spectral analysis on the emission from the whole western jet are in agreement with those presented in \citet{HZ09} and, for the first two detections, in \citet{Kaa03}.
The best-fit values of the photon index, $\Gamma$, are consistent with being constant through the period of the X-ray observing campaign, with the exception of the observation in September 2002 (ObsID 3807), when the spectrum is the steepest ($\Gamma=$2.15\er0.16). 
Note that the $\Gamma$ of the latter observation by \citet{HZ09} is flatter than ours, although still in agreement within the errors. This could be due to some differences in the analysis method, as they used background-subtracted rebinned spectra and applied the chi-squared statistic. 
The 0.3-8.0 keV flux is constantly decreasing, with a reduction of about a factor of 6 from the first to the last observation.  

We also tested the data with a model of thermal emission from a hot gas, as previously done for the first two observations of the two jets  \citep[see][]{Tom03,Kaa03}. A thermal bremsstrahlung model  \citep[\texttt{xsbremss} in Sherpa, He abundance is assumed to be 8.5\% of H by number, see][]{Kel75} gives an acceptable fit to the data. The best fit values of the temperature parameter range between 3.5\er0.9 keV and 6.0\er2.0 keV, but the large uncertainties on the measures do not allow us to conclude on a variation of the plasma temperature.

Finally, for each observation we measured the 0.3-8.0 keV flux produced in the tail region (see Sec. 3) assuming a power-law model with Galactic absorption (in Table \ref{t3spec}). 
In addition, the spectra of the tail region were also fitted with a thermal bremsstrahlung model to test the hypothesis that a fraction of the total emission, corresponding to the faintest emission, is produced by hot diffuse gas (see Sec. 5.2). 

\subsection{Radio Fluxes and Spectra}
The radio flux densities of the western jet in Table \ref{t1} were obtained from the ATCA multifrequency radio observations. When the radio jet was resolved, we measured the fluxes of the different components by fitting the radio image with multiple elliptical Gaussians. The quoted errors take into account the systematic error and the statistical error of the fit.  Given the limited number of observations at the best spatial resolution and their different sensitivity, we did not attempt to establish a correspondence among the components in the different observations. 

In Figure \ref{f10a} (upper panel), we plotted the measured radio fluxes at the four frequencies as a function of time. The time is measured in days starting from the date of the X-ray and radio flare of September 1998 \citep[MJD=51078,][]{Sob00,Han01,Han09}. Two phases can be distinguished in the light curves at 4.8 and 8.64 GHz: an initial plateau (January 16, 2002 to May 2, 2002, i.e. MJD=52291 to MJD=52397), which corresponds to a period of maximum and rather stable (or slowly decreasing) flux, and a following rapid decay  (May 22, 2002 to July 25, 2003, i.e. MJD=52416 to MJD=52845). A similar trend, although with less available data points, is also found in the lightcurves at 1.4 GHz and 2.5 GHz. The decay phase presents some degree of flux fluctuation. On September 17, 2002 (MJD=52534.2), we observed a re-brightening of the radio jet at 8.64 GHz, with the flux increasing and dropping by more than a factor 2 within a time interval of one month. 
Before January 2002 the western jet is detected only once, and for the first time, on February 9, 2001  (MJD=51950) at 8.64 GHz. The flux density, $F_{\rm 8.64GHz}=0.25$\er0.08 mJy, is about a factor of 10 lower than that observed in the plateau phase, indicating that in 2001 the radio flux is still rising. Thanks to this first detection, we can place a loose constraint of $\lesssim$340 days to the phase when the flux is rising.

 We fitted the decay phase at each frequency with a power-law model, $ f(t)=f(t_{\rm 0})\,(t/t_{\rm 0})^{-\beta}$, and an exponential model, $f(t)=f(t_{\rm 0})\,e^{-t/t_{\rm decay}}$, where ${f(t_{\rm 0})}$ is the flux at the time $t_{\rm 0}$ when the flux decay begins and $t_{\rm decay}$ is the 1/$e$-folding time. We assumed that the radio fluxes in the plateau phase correspond to the peak of the radio emission. We did not try to link the $t_{\rm 0}$ and referred to the ATCA constraints at each frequency. For both models, the beginning of the flux decay, $t_{\rm 0}$, is 1319 days after the 1998 outburst (May 2, 2002, i.e. MJD=52397, respectively) at 4.8 and 8.6 GHz. In the less sampled 1.4 and 2.5 GHz lightcurves, $t_{\rm 0}$ was set 1256 after the 1998 outburst (March 11, 2002, i.e. MJD=52334), although, as we verified, placing $t_{\rm 0}$ in the same date of the higher frequencies did not change our conclusions. We excluded from the fit the fluxes at the time of the re-brightening at 8.64 GHz.  The power-law model provides a poorer description of the datasets than the exponential decay at all four radio frequencies (see the fits to the 8.64 GHz data in Figure \ref{f12}). Within the same time interval, the timescales of the exponential decay of the two best sampled frequencies are significantly different. The 1/$e$-folding time is $t_{\rm decay}=$167\er5 days at 4.8 GHz and  $t_{\rm decay}=$127\er6 days at 8.64 GHz, whereas we obtained a $t_{\rm decay}$ of 125\er31 days and of 85\er26 days at 2.5 GHz and 1.4 GHz, respectively (keeping in mind the limited number of points at these frequencies). 

When fluxes in at least three observing frequencies were available, we measured the radio spectral index, $\alpha_{\rm r}$, by fitting the spectrum with the function $f_{\nu}\propto \nu^{\alpha_{\rm r}}$, where $f_{\rm \nu}$ is the flux density and $\nu$ is the frequency. We directly measured the spectral slope when only the fluxes in two frequencies were available. 
As an example, in Figure \ref{f11a} (upper panel) we show the radio spectra and the best-fit power-laws of observation17 (upper panel) and 18 (lower panel). The radio spectral indexes are listed in  Table \ref{t1}. The $\alpha_{\rm r}$ vary in time but it always remain negative, as expected for  optically thin synchrotron emission.
In Figure  \ref{f10a} (lower panel), the $\alpha_{\rm r}$ values are plotted as a function of the time. In the period corresponding to the lightcurve plateau, the radio spectral index changes within the interval $-0.8\lesssim \alpha_{\rm r} \lesssim-0.6$. The largest variation of $\alpha_{\rm r}$ is measured during the decay phase of the flux: there is an initial steepening trend until a minimum of $\alpha_{\rm r}\sim-$1.1 is reached. This trend breaks in September 2002, at the time of the flare at 8.64 GHz, when the less steep  radio spectrum ($\alpha_{\rm r}=-$0.43\er0.07) is measured.
After that,  there is not a clear trend, such as a progressive steepening of the spectrum that might be expected if radiative losses were responsible of the spectral evolution.

\subsection{Polarization}
Linearly polarized emission at $\geq$3$\sigma$ significance was detected at 4.8 GHz and 8.6 GHz when the western jet was the radio brightest. 
The fractional linear polarization (FP) is $\sim$9\% at 4.8 GHz between January and June 2002 (specifically, we measured FP=8.7\%\er0.9\%, 9.0\%\er3.5\% and 9.5\%\er3.0\% in observations 4,10 and 11, respectively, see Figure \ref{f11b}) and rapidly drops as the total radio flux decreases, FP$\lesssim$2\% in August 2002 (observation 16 in Figure \ref{f11b}). 
Measurements of the fraction of polarized flux are more difficult at 8.6 GHz because of the lower total flux, however we were able to measure a maximum FP=9\%\er1\% on April 8-9, 2002, which is drastically reduced (FP$\sim$2\%) in about 40 days (observation 11 in 2002, Figure \ref{f11c}, right panel). 
In the 8.6 GHz map of April 8-9, 2002 the jet is resolved and  the polarized flux appears concentrated in the leading edge region (see Figure \ref{f11c}, left panel). 

The electric vector polarization angle (EVPA) at 4.8 GHz is  EVPA(4.8)=84\dg\er6\dg~at the peak of the FP in January 2002, consistent with the EVPA(4.8)=78\dg\er5\dg~measured at the end of May 2002 (observation 11) and with the EVPA at 8.6 GHz FP peak, EVPA(8.6)=86\dg\er7\dg.  
The similar EVPAs at 4.8 GHz and 8.6 GHz suggests that there is no significant Faraday rotation effect, however the observations in the two bands are not simultaneous (and that N$\pi$ ambiguities might affect the measurements).  We used the publicly available maps of the Galactic Faraday depth to obtain an estimate of the Faraday  rotation \citep[][]{Opp12,Opp15}. The value of the Faraday rotation reported at the position of \xte~($-$102 rad m$^{-2}$) translates in a difference between the observed and intrinsic EVPAs of 23\dg~and 7\dg~at 4.8 and 8.6 GHz, respectively (and thus in a difference  between the EVPAs at the two frequencies of $\sim$16\dg). These estimates should be considered as upper limits, because the Faraday rotation was measured using galactic and extragalactic sources, thus at distances that are larger than \xte~\cite[and for a minimum angular scale of half-degree, see][for details]{Opp12}.
A EVPA(4.8)=66\dg\er4\dg  is measured at the beginning of May 2002 (observation 10, see Figure \ref{f11b}). This variation of the EVPA would thus occur in between two consistent measurements, on a $<$2 month timescale. A different EVPA, EVPA(4.8)=$-$39\dg\er8\dg, is also found in August 2002 (observation 16), however we note that at this epoch we have only a marginal detection of FP. For the same reason, it is difficult to define the reliability of the average EVPA at 8.6 GHz measured at the end of May 2002, EVPA(8.6)=$-$35\dg\er9\dg, which is significantly different from the measurement at 4.8 GHz in the same observation (see Figures \ref{f11b} and \ref{f11c}, respectively). In this latter observation, we observe that the polarized flux at 8.6 GHz appears to extend to the rear region, with respect to the peak of the total emission, differently from what is found in the other cases (with the exception of the observation 16 in August 2002, when the polarized emission appears shifted slightly behind the peak of the total emission).

All in all, we conclude that the electric vector is oriented parallel to the jet main axis, with indications of variations during the period of the  ATCA observations.

\subsection{X-ray Lightcurves}
The time variation of the 1 keV flux density of the western jet is shown in Figure \ref{f12}. We included the upper limits (at 90\% confidence limit) to the detection of the western jet in the \Cha observations of August and September 2000 (ObsID 1845 and ObsID 1846). These were obtained with the \texttt{srcflux} script by selecting a large rectangular region along the jet main axis, covering the region between the position of the western jet at its first detection and the position of XTE J1550$-$564. The upper limits obtained here ($F_{\rm 1keV}<1.6\times10^{-15}$ \ergcm~and $F_{\rm 1keV}<2.6\times10^{-15}$ \ergcm) are lower than those reported by \citet{Kaa03}, because the \texttt{srcflux} script accounts for the background contribution to the flux.  We also show the evolution in time of the 1 keV flux of the eastern jet, for a comparison between the two jets. The first flux point of the X-ray lightcurve of the eastern jet is from \cite{Tom03}, in addition we included the fluxes that we measured in the  \Cha~observations of September 2002 (ObsID 3807) and  July 2003 (ObsID 5190). The upper limits in the observations in May 2002 (ObsID 3672) and January 2003 (ObsID 4368) were obtained with the same procedure described for the western jet upper limits. The rectangular extraction region covered the area between the positions at the first and last detections of the X-ray eastern jet.
 
Both X-ray jets are caught during the phase of flux decay. However, using the upper limits we are able to place a time constraint of $\lesssim$550 days to the stage during which the western jet X-ray flux rose and peaked.
As done in the radio band, we fitted the X-ray light curves with an exponential decay curve and a power-law decay. The X-ray upper limits of the western jet set a lower limit to the beginning of its flux decay, $t_{\rm 0}>$720 days after the 1998 outburst (i.e. September 2000). 
 We made the hypothesis that the X-ray emission took some time to reach (at least) the maximum observed level, assuming an exponential rise of the flux, and we ended up with 850 days$\lesssim t_{\rm 0}\lesssim$1270 days (approximately between January 2001 and March 2002), with the upper bound coincident with the epoch of the first X-ray detection. Note that we did not assume any link between the $t_{\rm 0}$ of the X-ray lightcurve and that of the radio observations.  For $t_{\rm 0}=$850 days, both models provide acceptable fits to the X-ray data. The 1/$e$-folding time of the exponential decay is $t_{\rm decay}\sim$338\er14 days, consistent within the uncertainties with that of the eastern jet $t_{\rm decay}\sim$311\er37 days\footnote{Here, the $t_{\rm decay}$ are estimated in the observer's reference frame. Given the non-linearity of the motion of the two jets (see \ref{propmot}) and their interaction with the local ISM, which is likely to affect how their emission evolves, we did not attempt to apply the time correction between the approaching and receding jets.}. The best-fit index of the power-law is 2.00\er0.09. Such a value is significantly flatter than that reported by \citet{HZ09}, which was obtained setting $t_{\rm 0}$ at the time of the 1998 X-ray outburst, thus without taking into account the constrains of the upper limits. We note however that the results of their modeling are still compatible with the observational constraints (see Figure 5 in their paper).
The predicted 0.3-8 keV flux at $t_{\rm 0}$ is 7.2$\times$10$^{-13}$ \ergcm~for the exponential decay, and 4$\times$10$^{-8}$ \ergcm~for the power-law decay. The latter value, although large, is $\sim$10\% of the flux at the peak of the 1998 X-ray outburst.  
When $t_0\geq$900 days ($\approx$March 2001) is assumed, the fit using the power-law model significantly over predicts the flux at the time of the two last \Cha~observations.

\subsection{Jets' Radio and X-ray Positions \& Proper Motion}
\label{propmot}
The position and separation in the \Cha~observations of the large scale jets from \xte~were used  to constrain the models of the jet kinematics \citep{HZ09,SM12}.   
The intensive observational campaign in the radio band allows us to trace the jet proper motion with unprecedented accuracy. 
In Table \ref{t4radsep} we report the angular separation of the western radio jet with respect to XTE J1550$-$564. When available, we gave preference to the measurements at 8.64 GHz, and used the 4.8 GHz maps in the remaining cases.  If the radio jet was resolved, we used the position of its brightest component.
In the observations where the central compact source was not detected,  we referred to the radio position of XTE J1550$-$564 (RA(J2000)=15h 50m 58s.7 DEC(J2000)=$-$56$^\circ$ 28$^\prime$ 35.2\arcsec) reported in \cite{Cor01}.  The uncertainties on the angular separation take into account the statistical and systematic errors on the positions. 
The angular separation between XTE J1550$-$564 and each jet as a function of the time (the origin of the time is again the flare of September 1998)  is shown in Figure \ref{f13}. 
In X-rays, we used the centroid positions of the two jets obtained with \texttt{wavdetect} (see Tables \ref{t2} and \ref{t5}), except for the positions of the eastern jet in ObsID 3807 and 5190, which were measured with the centroid task of the ds9 analysis extensions (dax). Because of the jet faintness, we assigned to these two data points a larger uncertainty correspondent to the extension of the eastern jet emission in the X-ray image. Note that in each X-ray detection the position of the centroid and of the projected peak are consistent within the errors, therefore, for the purpose of studying the jet motion,  the two quantities are basically the same. 
The angular separation of the eastern jet measured in the ATCA observation of 2001 is also included, together with the radio position reported in \cite{Cor02}. \\
\indent
Following \cite{Kaa03}, we used a K-S test to quantify the shift of the peak position in the brightness profiles of the X-ray western jet in the five \Cha~observations.  In Table \ref{t2}, we report the shift of the peak  with respect to the position at the first X-ray detection. A significant offset of the peak with respect to its initial position is found in ObsID 3672 \cite[0.52\arcsec\er0.12\arcsec, as reported in][]{Kaa03} and 3807 (0.70\arcsec\er0.12\arcsec), with probabilities of a zero offset $<10^{-11}$. In the last two observations, the K-S test seems instead to underestimate the actual offset of the peak (1.75\arcsec\er0.35\arcsec for the last observation, using the bin size as the error on the peak position).  
However, in these two cases we obtained low probabilities that the two samples are drawn from the same distribution (see Sec. 4.1). Most likely, this result is a consequence of the strong evolution of the morphology of the jet. \\ 
\indent
We calculated the speed of the jet in two ways: 1- the average jet's proper motion with respect to \xte~(v$_{\rm app.,xte}$); 2- the advancing velocity of the jet with respect to the position of  its first detection (v$_{\rm app.,3448}$, v$_{\rm app.,obs1/01}$ for the X-ray and radio data, respectively). We separately considered the positions of the jet in the radio and X-ray observations.\\
\indent
We used the angular separation of the western jet  from XTE J1550$-$564 in the radio and X-ray observations to calculate the jets' proper motion, v$_{\rm app.,xte}$. The average proper motions at each epoch, measured with respect to the supposed time of formation of the jets in September 1998 (MJD=51078), are reported in Tables \ref{t2}, \ref{t4radsep} and shown in Figure \ref{f13} (lower panel). 
The average proper motions obtained from the radio observations confirm the results of the X-ray studies \citep{Cor02,Tom03,Kaa03}. In February 2001, we measured a proper motion of the western radio jet of 23.7\er0.6 mas day$^ {-1}$, while the value that we derived from the first X-ray detection \citep[17.9\er0.4 mas day$^ {-1}$,][]{Kaa03} is in perfect agreement with that from the coeval radio observations (see obs 8 and 9 in Table \ref{t4radsep}). Over the following two years, the jet decelerates until reaching the final average speed $\sim$13.0\er0.3 mas day$^ {-1}$.
 For a comparison, we calculated the v$_{\rm app.,xte}$ of the eastern X-ray jet (in Table \ref{t5}). The v$_{\rm app.,xte}$ of the eastern X-ray jet was 34.0\er0.7 mas day$^ {-1}$ at the time of the first X-ray detection in June 2000 (ObsID 679) and dropped to 17.9\er0.8 mas day$^ {-1}$  in October 2003.\\
\indent 
In X-rays, we used the offset of the peak of the western jet with respect to its position in the first detection to calculate v$_{\rm app.,3448}$ (see Table \ref{t2}), and did the same for the radio data (v$_{\rm app.,obs1/01}$ in Table \ref{t4radsep}). In October 2003 we measured v$_{\rm app.,3448}$=3\er1 mas day$^ {-1}$, in agreement with the velocity from the radio observations, v$_{\rm app.,obs1/01}$=3.1\er1.1 mas day$^ {-1}$. These final values clearly show that within $\sim$2 years the western jet underwent a drastic deceleration, with a decrease up to a factor $\sim$8 with respect to the initial speed. The amplitude of the deceleration of the western jet appears larger than that of the eastern jet: the final advancing velocity of the eastern jet detection is v$_{\rm app.,679}$=10\er1 mas day$^ {-1}$, indicating a decrease of a factor $\sim3.5$ from its initial velocity in 2000.

\section{Discussion} 
\label{discussion}
We have investigated the morphology, the motion and the emission of the western jet of \xte~making use of X-ray and radio observations. The campaign of observations began with the discovery of the large scale jets \citep{Cor02} and covered a $\sim$1.5 year period.
The analysis of the radio and X-ray data showed the remarkable evolution of the jet morphology, the chromatic decay of the broadband emission and confirmed the jet deceleration. In the following, we discuss the inferences from these results on the structure of the jet, the nature of the emission and the mechanism that produced it.

\subsection{A receding tail} 
\label{tail}
The formation of a tail in the X-ray structure of the jet is a major, surprising result of the morphological analysis. 
 The tail seems not to participate in the advancing motion of the jet main body, rather, its endpoint appears to shift backwards in the direction of the central source (see Sec. 4.1 and Figure \ref{f4}). 
We used the shift of the tail endpoint between the first and last X-ray observations  (2.25\arcsec\er0.6\arcsec) to measure the proper motion of the tail $\mu_{\rm app,tail}=-$4\er1 mas day$^{-1}$ (the motion is defined along the jet main axis and the negative sign indicates the receding direction). 
  Its corresponding apparent velocity at the source distance of 4.4\er0.6 kpc \citep{Oro11} is $v_{\rm app,tail}\sim$($-$0.10\er0.02)$c$. 
 For a comparison, the apparent velocity of the peak between the first and last detections is (0.07\er0.02)$c$ ($\mu_{\rm app,peak}=$3\er1 mas day$^{-1}$). 
  The emerging picture is that of a rear region that is rapidly expanding backwards while the front region is progressively decelerating (see Table \ref{t2}).
  
The measured $v_{\rm app,tail}$ is large for a plasma that is back flowing after the impact with the ISM. In addition, there is no clear evidence of radio lobes associated with the  large scale jets of \xte.
On the other hand, the low surface brightness of the microquasars' radio lobes hampers their direct detection, and leaves the best clues of their presence to the fact that the jets travel unseen through a low-density medium \cite[see][for a discussion]{Hei14}, as proposed for \xte~and H1743$-$322 \citep{HZ09,SM12,SMR12}. 

The tail could be formed by plasma that is still within the jet. Here, we discuss two possible scenarios, which could account for the observed feature.
If the X-ray emission is produced by particles accelerated by a reverse shock \citep{Wan03,HZ09}, the tail could be the signature of the shock moving backwards through the jet's plasma. The propagation speed of the shock with respect to the plasma, and in the reference frame of the observer, depends on the conditions (e.g., the density of the plasma and ISM, the jet bulk motion) in which the shock develops.
Observational evidences of a reverse shock moving backward through a plasma envelope has been reported for supernova remnants (SNRs): sensitive X-ray observations have revealed a decrease in time of the reverse shock radius in Tycho's SNR, pointing to a reverse shock which propagates inward with  a Mach number $\sim$1000 \citep{Yam14}. 
We note that the shocks observed in the supernova remnants are non relativistic, while the shocks  in the western jet of \xte~may still be in a (mildly) relativistic regime.  In principle, the deceleration motion of the western jet of \xte~and the density jump at the cavity wall are favorable conditions to the development of a reverse shock propagating with a larger speed than the bulk advancing speed of the jet, however numerical simulations are necessary to properly define the dynamics of such a system and validate this hypothesis.

Alternatively, it is possible that the emission is  produced via internal shocks, which develop in the collision among discrete shells of plasma, coming from the central source at different times and with different velocities. This mechanism is typically considered for the compact jets of XRBs \citep{KSS00,JFK10,Mal13,Mal14} and AGNs \citep{Spa01}, but it could still play a role on large scales. For example, in the nearby radio galaxy 3C~264, a collision between two knots of the jet has been imaged at kiloparsec distances from the central AGN  by collecting 20 years of optical observations with the Hubble Space Telescope \citep[][]{Mey15}. In the internal shock scenario, the tail could be formed when late-arriving shells collide against the plasma that has been decelerated by the ISM. Thus, the receding motion would be an apparent effect. 
 Such a hypothesis requires that the ejecta are produced from the central object over a longer time period than the impulsive event observed in 1998 \citep[with a duration of $\lesssim$10 days][]{Wil98,Rem98}. Interestingly, after the 1998 outburst, \xte~became active again in April 2000, with an increased of the emission observed in the X-ray, radio and optical-IR bands  \citep{Smi00,Cor01,Jai01}. Analysis of the broadband spectrum supported a significant, if not dominant, contribution of the radio jet to the observed optical and X-ray emission during the (hard-state) decline of the outburst \citep{Rus10}. 
It is unlikely that collisions between material ejected during the first and second outbursts are responsible for the whole observed emission. In fact, the first detections of the two radio jets happened in June 2000 and February 2001, too early for the ejecta of the April 2000 event to cover the $\approx$0.5-0.6 pc distance. However, plasma bullets leaving the central object in 2000 at a speed similar to that measured initially for the western jet ($\sim$24 mas day$^{-1}$, see Table \ref{t4radsep}) could have reached the location of the western jet by the second half of year 2002. 
In this framework, the re-brightening that we observed at 8.6 GHz in September 2002 (observation 17) could be interpreted as an additional boost of emission due to particles accelerated in late time collisions. The flattening of the radio spectral index could be then explained with the injection of newly accelerated particles in the shell collisions. 
The flare in September 2002 is not observed in X-rays. We cannot exclude that an X-ray flare occurred in between two \Cha~observations. Alternatively, we can make the hypothesis that, in this second event, electrons were not accelerated up to the energies necessary to produce (additional) synchrotron X-ray emission.

\subsection{Broadband emission of the western jet}
The broadband spectra of the western jet were obtained using the radio and X-ray (1 keV) total fluxes of the observations which were close in time (note that, differently from the morphological analysis, in this case it was not necessary to select the ATCA observations at the best angular resolution).

 A simple power law is used to model the radio to X-ray SED (Figure \ref{f14}). The broadband spectral index $\alpha_{\rm rx}$ does not vary significantly in the four SEDs ($\alpha_{\rm rx}\sim-0.62$) and is (broadly) consistent with the correspondent $\alpha_{\rm r}$ in four over five cases. Instead, $\alpha_{\rm rx}$ is always, to a variable extent, less steep than the measured X-ray spectral indices (for uncertainties at the 90\% confidence level). We conclude that, although there are indications for a different spectral component, the majority of the SEDs are consistent with a simple power law shape. The notable exception is the broadband spectrum obtained from the observations in September 2002 (ObsID 3087 and obs17 in X-ray and radio, respectively, in Figure \ref{f14}). The radio observation corresponds to the re-brightening of the 8.64 GHz flux, when we measured the less steep radio spectral index ($\alpha_{\rm r}=-$0.43\er0.07). On the contrary, at the same epoch in X-rays we registered the steepest spectral index ($\alpha_{\rm X}\equiv -(\Gamma-1)=-$1.15$^{+0.16}_{-0.14}$). An extrapolation of the radio(/X-ray) spectrum to the higher(/lower) energies overestimates the observed X-ray(/radio) flux, even taking into account the uncertainties on the spectral parameters (see Figure \ref{f14}).  
 
 If we model the broadband spectrum with a broken power law and set the lower and higher spectral indexes equal to the best fit $\alpha_{\rm r}$ and $\alpha_{\rm X}$, respectively, the spectral break is at, approximately, a few $\sim$10$^{15}$ Hz. 
In compact jets, spectral breaks, marking the transition from optically thick to optically thin synchrotron emission, are directly observed or inferred within a large range of frequencies \citep[starting at 10$^{12}$ Hz to $>$10$^{14}$ Hz, see][and references therein]{CF02,Cor09,Gan11,Rah11,Rus13}.  Re-flares of the synchrotron emission associated with flattening of the radio-optical spectrum may occur in correspondence of an increase in the density of the particles accelerated by the shock \citep[][]{Broc05,Rus12,Cor13}.
On the other hand, we recall that the X-ray and radio (8.64 GHz) morphologies of the jet in this epoch are also different. Therefore, we cannot exclude that (part of) the synchrotron radio emission is produced by a second population of electrons, which is accelerated in a different site of the jet with respect to the high-energy particles responsible of the X-ray emission.

\subsubsection{X-ray emission mechanism}
The non-thermal, synchrotron origin of the radio emission is firmly established by the relatively high fraction of polarized radio emission.  The values of $\alpha_{\rm r}$ are consistent with the emission being optically thin (i.e. $\alpha_{\rm r}\lesssim$0) during all radio observations.
A synchrotron mechanism was proposed as the most likely mechanism producing the X-ray emission of both the large scale jets of XTE J1550$-$564 based on the radio-to-X-ray spectra and the similar morphologies in the two bands \citep[see][for the eastern jet]{Cor02,Tom03}. 
Synchrotron X-ray radiation is also likely in the case of the large scale jets of the microquasar H1743$-$322 \citep{Cor05,HZ09}.
On the other hand, in the X-ray binary SS 433 the detection of emission lines in the X-ray spectrum of the jets points to a contribution of thermal bremsstrahlung emission \citep{Mig02}.  \citet{Tom03} showed that a thermal bremsstrahlung origin is disfavored in the case of the X-ray emission of the  eastern jet of \xte. In fact, a large ($\gtrsim$5$\times$10$^{28}$ g) gas mass is necessary to produce the observed X-ray emission via bremsstrahlung, implying either  long ($>$1000 yrs) timescales to be accumulated through the central accretion, or an extremely efficient mass entrainment along the jet path. Nevertheless, it is still possible that X-ray thermal emission is produced by the ISM heated up in the collision with the jet \citep{Tom03}.
 
A thermal origin of the X-ray emission of the western jet, or of a fraction of it, may account for the differences between the radio and X-ray morphologies.  In fact, although the radio emission appears also extended, the presence of a radio counterpart to the X-ray tail is not clearly established by the ATCA observations (see Figure \ref{f9}).
Following \citet{Tom03}, we calculated the mass required to obtain the total X-ray flux of the western jet, and the X-ray flux of the tail region under the assumption of a thermal origin.
 The jet's half opening angle, $\Theta$, was set to 1\dg, in rough agreement with the extension of the X-ray emission in the direction perpendicular to the main axis \citep[see Sec 4.1 and ][]{Kaa03}. For a conical section shape of the emitting region, the estimated total volume of the jet in the last X-ray observation is $\sim$10$^{51}$ cm$^{3}$ at the source distance of 4.4 kpc. Indeed, one should consider that this estimate is subject to uncertainties due to e.g. our ignorance of the jet's actual depth, the errors on the source's distance and the inclination angle of the jet.
The estimated minimum masses of the whole jet, $\gtrsim$2$\times$10$^{29}$ g, and of the tail, $\gtrsim$9$\times$10$^{28}$ g, raise the same objections discussed by \citet{Tom03}. 
A rough estimate of the mass of the surrounding medium that is entrained and swept up by the shock is $m_{\rm sw}=\Theta^2 m_{\rm p} n \pi R^3/3$, where $R$ is the jet distance from the central object, $m_{\rm p}$ is the proton mass and $n$ is the ISM particle density \citep[see][]{HZ09,SM12}. Assuming $R\approx$0.6 pc (for the maximum jet extension and a distance of the source of 4.4 kpc), the maximum $m_{\rm sw}$ is $\lesssim$6$\times$10$^{27}$ g if $n=1$ cm$^{-3}$ and decreases for smaller values of $n$ ($\leq 0.1$), which are assumed inside the cavity \citep{HZ09,SM12}, hence enlarging the discrepancy with the masses inferred from the X-ray spectra.
We also note that the thermal X-ray emission produced by the ISM should be found in front of the jet, rather in its rear region.

If the X-ray emission of the tail is non-thermal, the non-detection at radio frequencies could be due to the limited sensitivity of the ATCA observations. We checked this possibility for the the third (ObsID 3807) and last (ObsID 5190) detections of the western X-ray jet, when the tail is the most visible, and their closest-in-time radio observations, obs17 and obs24. 
The 3$\sigma$ upper limit at the X-ray tail position ($\lesssim$0.39 mJy) obtained from the 8.64 GHz map of obs17 implies a radio to X-ray spectral index $\alpha_{\rm rx}\gtrsim -$0.7 in order for the extrapolation of the X-ray emission not to be detected in the radio band, if the radio-X-ray SED of the tail is a simple power-law. This is in slight tension with the uncertainty interval measured for the X-ray spectral index of the tail emission (see Table \ref{t3spec}). However, we note that also the broadband SED of the total emission at this epoch is best fitted by a broken power law. In the case of the last X-ray observation, a $\alpha_{\rm rx}> -$0.6, which is within the uncertainty interval of the $\alpha_{\rm X}$ of the tail, would correspond to radio fluxes  compatible with the 4.8 GHz flux of the component labelled G2 of observation 24 (in Table \ref{t1}), which is the most spatially coincident with the X-ray tail. Therefore, we cannot exclude that a radio emission due to synchrotron mechanism is present but too faint to be detected in the ATCA observations.

\subsubsection{Constraints on the particle acceleration process from the broadband emission}
The model proposed  to explain the emission of the large scale jets of \xte~is borrowed from the theory of  gamma-ray burst (GRB) afterglows. The afterglow is produced by particles that are accelerated by external shocks formed in the interaction of the GRB ejecta with the circumburst medium.
In the standard scenario, the X-ray emission is typically ascribed to the forward shock, which  propagates into the ambient medium, whereas the reverse shock, which travels through the GRB plasma, is responsible for the emission at lower, optical to radio, energies (Meszaros \& Rees 1997, Sari \& Piran 1999). 
 It was proposed by \citet{Wan03} that the same processes should be active in the jets of \xte, once rescaled for the typical bulk motions and energetics of microquasars. The authors
showed that the X-ray emission of the eastern jet faded faster than what expected for emission originating from the forward shock, which is supposed to continuously heat the ISM. Conversely, the reverse shock operates  only once, leaving then the shocked gas to expand and adiabatically cool. The same scenario was considered for the western jet \citep{HZ09,SM12}. 
Notwithstanding the constraints on t$_{\rm 0}$, our revised decay indices of the X-ray fluxes are still too steep ($\gtrsim$2) with respect to the model predictions for the forward shock \citep[1.2, assuming that the energy distribution of the accelerated particles has a power-law shape with an index p=2.2,][]{Wan03}.

The scenario of a reverse shock that is dominant in X-rays has been considered to explain some GRBs' afterglows \citep[see ][for a review]{Gao15}.
\citet{GDM07} have shown that the X-ray emission from a reverse shock may be dominant during the early afterglow phase provided that: 1- the bulk motion of the material ejected during the last stages of the source activity is $\Gamma_{\rm bulk}<$10 and 2- a large part of the shock-dissipated energy is transferred to a small fraction of the electron population \citep[see also ][]{UB07}. Interestingly, the bulk motions of the late ejecta  required by these models  are similar to those  that we expect in the jets of the microquasars. At the same time, only a small fraction of the total shock energy should be transferred to the particles ($e_e\lesssim$0.1) and magnetic field ($e_B\lesssim$0.0004) accelerated in the forward shock in order for its emission to be suppressed, while equipartition conditions are assumed for the reverse shock \citep[][]{Wan03}.

The fact that a power law model does not reproduce the X-ray and radio light curves of the western jet is possibly indicative of a second radiative component emerging at later times.
\citet{Wan03} argued that a late time flattening of the X-ray lightcurve could be the signature of the forward shock emission, which is decaying at a slower rate and becomes visible over the reverse shock emission. Alternatively, the model of Genet, Daigne \& Mochkovitch (2007) predicts a second phase of shallow decay of the reverse shock emission.

The reverse-shock model of \citet{HZ09}  was built on the X-ray data. Our radio campaign allows for an extension of the investigation to the low energies.
 In the radio domain, the evolution in time of  the early afterglow has not been extensively investigated because of the difficulties in obtaining radio observations at such early stages \citep[e.g.,][]{Las13,Per14,Las15}. Recently, early time observations of the afterglow of GRB~130427A with the Arcminute Microkelvin Imager (AMI) have captured the rise and rapid decline of the radio emission, most likely originating in the GRB reverse shock \citep{And14}.
 The decay of the radio flux of \xte's western jet  is more rapid than what is typically expected by the theory \citep[see e.g. the simulations\footnote{http://cosmo.nyu.edu/afterglowlibrary/index.html} presented by][]{van11}, however this result may be significantly modified by the different conditions of the ejecta in GRBs and microquasars, such as the kinetic power, opening and inclination angles. 
 
 Dominant radiative losses should induce a progressive steepening of the spectrum at the high energies. 
Except for one epoch, in September 2002 (Figure \ref{f14}), overall, the X-ray spectral index did not significantly change. The faster decay of the radio emission with respect to the X-rays is also in contrast with dominant radiative cooling. Negligible radiative losses are in line with the estimated synchrotron cooling times of the electrons producing the X-ray emission in the jets \citep[$\gtrsim$6 yrs,][]{Tom03}. 
On the other hand, in an uniformly expanding, homogenous plasma we would expect an achromatic decrease of the broadband spectrum and similar decays indexes at all frequencies.
In the framework of dominant adiabatic cooling, the different $t_{\rm decay}$ of the radio and X-ray emission could be due to the inhomogeneity of the plasma in the jet. 
In fact, the angular resolution of our observations cannot distinguish the presence of clumps of plasma with variable density and different expanding velocities. In addition, multiple acceleration episodes may also shape the light curves (as the re-flare at the 8.6 GHz). 

We investigated whether the X-ray tail may be a reason of the different decay times. To this aim, we measured $t_{\rm decay}$ of the X-ray emission coming from the front region (excluding thus the tail emission). In this way, the $t_{\rm decay}$ of the X-rays is reduced to 240\er8 days. If we count the first X-ray observation as part of the plateau phase, rather than of the decay, and exclude it from the fit, we obtain $t_{\rm decay}=$214\er8 days. Indeed, this shows the importance of having equally sensitive and sampled observations for a fully consistent comparison among the observing bands. Nevertheless, the minimum $t_{\rm decay}$ in X-rays is still longer than the $t_{\rm decay}=$167\er5 days measured at 4.8 GHz.

It is interesting to note that \xte's jet is not the only case where the emission seems to decay faster at the lower energies. In fact, on shorter time scales, the large scale jets of H~1743$-$322 also display a steeper lightcurve in the radio band than in X-rays \citep[see Figure 2 in ][]{Cor05}. 
The extended ($\sim$6$^\prime$) X-ray emission associated with the black hole candidate 4U~1755$-$33 was interpreted as fossil radiation of the jet, after that the central source turned off \citep{AW03,Kaa06}. 
As for \xte~and H~1743$-$322, the X-ray emission was ascribed to the synchrotron mechanism, however in this case the sensitivity of the ATCA observations prevented even the detection of the putative radio counterpart \citep{Kaa06}. 

If the radio emission is due to the reverse shock, the orientation and intensity of the linearly polarized emission are related to the magnetic field of the ejecta after the compression and amplification due to the passage of the shock. In an optically thin plasma, as the steep $\alpha_{\rm r}$ indicates for the western jet, the electric vector should be perpendicular to the magnetic field  \citep{GS69,Lon94}. An EVPA that is parallel to the jet axis, as found in the western jet of \xte, would thus imply a B field that is perpendicular to the jet axis, as expected in the case of shock compression \citep{Lai80}. 
Similar configurations have been found in a handful of black hole binaries \citep[e.g. GX 339$-$4, GRO J1655$-$40, MAXI J1836$-$194][ respectively]{Cor00,Han00,Rus15} and in the relativistic jet of the neutron star X-ray binary Circinus X-1 \citep{Tud08}, while deep Very Large Array images of the jets of SS 433 have revealed a magnetic field that is aligned parallel to the local velocity vector \citep[see ][]{MJ08b}.

The episodic rotation of the EVPA, which we possibly detected (see Sec. 4.5) in April to May 2002,  could have different causes \citep[see the discussion in][ and their Figure 4]{Cur14}: the lateral expansion of the plasma;  an helical magnetic field; the illumination of the magnetic field behind the shock by back-flowing particles \citep{Dre87} .  At smaller linear scales, in the low-mass XRB Swift J1745$-$26, a rotation of the EVPA was observed  to accompany the evolution from a compact, self-absorbed jet to an optically thin plasma \citep{Cur14} and might indicate a variation in the geometry of the B field \citep[see also GRO J1655$-$40, GRS 1915$+$105 and XTE J1908$+$094,][ respectively]{Han00,Fen02,Cur15}. Large rotations of the EVPA were observed also in the resolved components of the milli-arcsecond jet of GRS 1915$+$105 \citep{Fen99}. In this case, the associated rapid decrease of the fractional polarization was explained as due to the increasing randomization of the B field within the discrete ejecta.  The drop in the intensity of the polarized emission in the western jet could be related to the structure of the jet becoming more complex. Given the existence of a preferential direction of expansion of the jet, we can speculate that the changing morphology is affecting more the ordered B field with respect to the total one.

\subsection{Jet motion}
The radio observations allowed us to better track the motion of the western jet. In Figure \ref{f13}, we show the angular separation of the western jet from \xte~as measured in the radio and X-ray bands. The refined trajectory is in substantial agreement with the trajectory defined by the X-ray data (Figure \ref{f13}, upper panel), and thus gives support to the dynamical description provided by the models of the two groups \citep{HZ09,SM12}. 

The first detection of each jet happened at radio frequencies: in 2000 for the eastern jet and in in February 2001 for the western jet (see Figure \ref{f11aa}). Most likely, both jets were caught at the beginning of the deceleration phase. The last detection of the ejecta is instead in X-rays. 
 It is interesting to note that the first detection of the western radio jet in February 2001 is consistent with the location of the cavity wall. Nevertheless, as noted by \citet{SM12}, it is likely that the single density jump assumed in their model is a relatively simplified description of the real ISM density profile. In fact, the earliest detections of the eastern radio jet happened when the jet was still traveling within the putative cavity, and this would be indicative of jet interactions with a clumpy medium. Finally, we note that the measured projected distance of the eastern jet in the last X-ray detection appears larger than the model predicted one \citep[see Figure 2 in ][]{SM12}, suggesting that the observed radiating material is moving at a slightly larger velocity than that in the previous detection. 
  One possibility is that, following the interaction with the ISM, the eastern jet underwent partial fragmentation.  In this case, the emission in the last X-ray observation could come from the debris that pierced through the  ISM, whereas the rest of the jet stalled at the cavity's boundaries. This scenario implies indeed some fine-tuning as the trajectory of the surviving X-ray component is basically unaltered.
 Alternatively in the framework of the colliding shell model, re-acceleration of the ejecta could be the consequence of late-time, ``billiard-ball'' collisions with new shells of plasma that are approaching at relativistic velocities (see Section \ref{tail}). 

\section{Summary}
\label{summ}
The microquasar \xte~represents one of the best observed and most investigated case of jet-ISM interactions.
Here, we presented a study based on an unique set of radio and X-ray observations, performed  between 2001 and 2003. 
We probed the inner structure of the large scale western jet and studied in details its properties and evolution over $\sim$2 years. Our analysis provides critical tests of the dynamical and radiative models proposed by different groups, and raises challenges for future theoretical work. 

We spatially resolved the complex inner structure of the X-ray jet. A main result of the analysis  is the discovery of the formation in the X-ray jet of a tail, which appears to develop backwards in the direction of \xte. The apparent velocity of the receding tail over 1.5 yrs of \Cha~observations is $v_{\rm app,tail}\sim$($-$0.10\er0.02)$c$, which is comparable with, if not larger than, in absolute value, the advancing motion of the peak emission.
In the framework of a reverse-shock dominated X-ray emission, as proposed by \citet{HZ09}, the receding tail may be the observational signature of the passage of the shock through the jet plasma layers. Instead, if the emission is produced by particles accelerated via internal shocks, the tail could be formed by collisions among late-time arriving plasma shells.

The detections of the jet in the radio band improved the mapping of the jet motion and confirmed the jet deceleration.  From the initial average velocity $v_{\rm app.,xte}=$23.7\er0.6 mas days$^{-1}$ in 2001, the jet reached the final value of $v_{\rm app.,xte}=$13.3\er0.5 mas days$^{-1}$ at the time of its last detection at 4.8 GHz in 2003, with a decrease of a factor of $\sim$8 from the first detection to the last one ($v_{\rm app.,obs1/01}=$3.1\er1.1 mas days$^{-1}$). The western jet was detected for the first time in February 2001. Interestingly, the first detection happened at the location where the dynamical models placed a jump in the ISM density profile \citep{HZ09,SM12}. Most relevant, in 2001, the radio flux was still in the rising phase, suggesting that the process of particle acceleration and radiative dissipation had just been triggered. Confirmation of this picture comes also from the fact that the jet was still undetected in X-rays in August/September 2000 and appeared for the first time, already in the decaying flux regime, in March 2002. Therefore, although other, most-likely minor episodes of energy dissipation may have occurred at shorter distances from \xte, the ATCA observations support the proposed scenario of a jet that is traveling through a low-density ISM cavity, undergoing little energy dissipation until the moment of the interaction with the denser medium. 

The multi-wavelength emission of the jet is not of straightforward interpretation. The radio and X-ray emission are most likely of synchrotron origin: the SED obtained from coeval radio and X-ray observations are consistent with a single spectral component in all cases but one. The radio emission is also spatially extended: we showed that a radio tail could also exist but remain unobserved due to the sensitivity limits of the ATCA observations. On the other hand, we observed variations in time of the SED: the SED of the western jet in September 2002 clearly deviates from a simple power law shape, with a spectral break that falls at $\sim$10$^{15}$ Hz frequencies. The re-brightening of the radio emission, together with a flattening of its spectral index ($\alpha_{\rm r}=-$0.43\er0.07),  which are observed at that time, could be due to the injection of freshly accelerated radiating electrons.  
The fact that the X-ray spectrum in September 2002 is the steepest and that the radio and X-ray morphologies of the jet appear the most markedly different, may imply different sites of production of the radio and X-ray emission. 

During the first half of 2002 we measured a significant fraction (up to $\sim$9\%) of linearly polarized emission at 4.8 GHz and 8.64 GHz, which rapidly drops as the radio flux decreases.  
The electric vector of the polarized emission is parallel to the jet main axis. If the synchrotron radio emission is produced by the jet plasma, the polarized flux indicates that the magnetic field is oriented perpendicularly to the main jet axis, as expected for a post-shock, compressed B field. {The decrease of the FP and} the  variation of the EVPA, that we possibly observed on month-timescales, could relate to changes in the local magnetic field structure (e.g. compression, turbulence) in consequence of the interaction of the ejecta with the ISM.

A challenge for the radiative models is the observed chromatic decay of the multi-band emission. Differently from what expected when adiabatic and/or radiative losses are dominant, the decay of the flux of the western jet was more rapid at low energies than at high energies. In addition, we showed that, if all the observational limits are taken into account, the decay of the radio and X-ray fluxes cannot be reproduced by a single power law model. A possible explanation is the emergence at late times of another emitting component (such as the forward shock emission or emission triggered by the collision among plasma shells). This finding, and the detection of the eastern jet in the last \Cha observation, highlight the importance of late-time, deep observations in order to follow the evolution of the emission of these large scale transient jets, and provide us with guidelines for the monitoring of future similar events.

\section*{Acknowledgements}
We thank the anonymous referee for a thorough reading and insightful comments, which helped to improve this work.
G. M. and S. C. acknowledge funding support from the UnivEarthS Labex program of
Sorbonne Paris Cit\'e (ANR-10-LABX-0023 and ANR-11-IDEX-0005-02) and from the French Research National Agency: CHAOS project ANR-12-BS05-0009.
P.K. acknowledges support from Chandra grant GO4-5038X.
G.M. thanks A. Siemiginowska and H.M. Guenther for precious support in performing Sherpa and MARX simulations. G.M. thanks F. Daigne, J. Steiner, P. Varniere and A. Bracco for useful discussions.
This research has made use of the software provided by the Chandra X-ray Center (CXC) in the application packages CIAO, ChIPS, and Sherpa.
This research has made use of TOPCAT\footnote{http://www.star.bris.ac.uk/$\sim$mbt/topcat/} \citep{Tay05} for the preparation and manipulation of the tabular data.






\begin{table*}
 \centering
 \begin{minipage}{166mm}
\caption{Western Jet -- ATCA Observations}
\label{t1}
 \begin{tabular}{@{}lccccccc@{}}
\hline
\hline
Obs     & Date \&           & MJD                     & Flux$_{\rm 8.6GHz}$            & Flux$_{\rm 4.8GHz}$          & Flux$_{\rm 2.5GHz}$            & Flux$_{\rm 1.4GHz}$            & $\alpha_{\rm r}$   \\
           & array config.   & (\& obs. length)     & (\& rms)                                & (\& rms)                               & (\& rms)                                 & (\& rms)                                 & \\
           &                        &(days, hours)         &(mJy, $\mu$Jy bm$^{-1}$)    &(mJy, $\mu$Jy bm$^{-1}$)    &(mJy, $\mu$Jy bm$^{-1}$)     &(mJy, $\mu$Jy bm$^{-1}$)     &    \\
(1)      &(2)                    &(3)                         &(4)                                          &(5)                                         &(6)                                          &(7)                                          &(8)\\
\hline
\\
\multicolumn{8}{c}{\bf Year 1999}\\
1/99         &03/11/99           &51248.68            &$<$0.18               &$<$0.3          & --                             & --                           &--\\
                &750D                &(2.04)                  &(60)                      &(100)                              &--                              &--                            &--\\
\\
\multicolumn{8}{c}{\bf Year 2000}\\
1/00         &04/30/00           &51664.61            &$<$0.18                   &$<$0.39        & --                             & --                           &--    \\
                &750D                &(2.49)                  &(60)                             &(130)                              &--                              &--                            &--\\
2/00         &05/06/00           &51670.53            &$<$0.24                    &$<$0.15         & --                             & --                           &--    \\
                &750D              &(1.97)                    &(80)                             &(50)                              &--                              &--                            &--\\
3/00         &06/01/00           &51697.15            &$<$0.12                  &$<$0.15         & --                             & --                           &--    \\
                &6D                 &(1.74)                     &(40)                             &(50)                              &--                              &--                            &--\\
\\
\multicolumn{8}{c}{\bf Year 2001}\\
1/01         &02/09/01      &51949.99                & 0.25\er0.09          &$<$0.18              & --                             & --                           &--\\
                &375              &(7.26)                     &(51)                        &(60)                              &--                              &--                            &--\\
2/01         &02/20/01      &51960.70               &$<$0.14                &--                              &--                                        &--                        &--\\
                &375             &(5.06)                      &(37)                             &--                              &--                              &--                            &--\\
3/01         &02/24/01      &51964.96               &$<$0.24               &--                              &--                                        &--                        &--\\
                &375             &(1.50)                     &(80)                             &--                              &--                              &--                            &--\\
\\
\multicolumn{8}{c}{\bf Year 2002}\\
1         & 01/16/02   &52290.86                      & 2.46\er0.09          & 3.76\er0.10             & --                             & --                           &$-$0.72\er0.08\\
           &750A         &(3.42)                             &(50)                          &(48)                              &--                              &--                            &--\\
2         &01/18/02    &52292.86                     & 2.45\er0.08           & 3.92\er0.10             & --                             & --                           &$-$0.80\er0.08\\
           &750A         &(7.08)                            &(50)                           &(50)                              &--                              &--                            &--\\
3         &01/29/02    &52303.72                     & 2.28\er0.10 (T)          & 3.64\er0.09        & --                              & --                           &$-$0.80\er0.09\\
           &6A              &(5.59)                           & 1.80\er0.15 (G1)          &(45)                              &--                                        &--                        &--\\
           &--                  &--                                &0.48\er0.15 (G2)           &--                              & --                                       &--                        &--\\
           &--                &--                                  &(50)                              &--                              & --                                       &--                        &--\\                    
4         &02/01/02    &52306.90                      & 2.18\er0.10 (T)          & 3.43\er0.11           & 5.49\er0.27              & 8.33\er0.43           &$-0.72\pm0.03$\\
           &6B              &(4.62)                           & 1.89\er0.14 (G1)        &(50)                       & (150)                                & (140)                      &--\\  
           &--                  & --                               &  0.29\er0.14 (G2)       &--                           & --                                & --                             &--\\  
           &--                &--                                 &(50)                              &--                           & --                                & --                             &--\\    
5         &02/14/02    &52319.73                    & 2.30\er0.09          & 3.63\er0.08           & --                             & --                           &$-$0.78\er0.08\\
           &1.5A          &(3.85)                           &(50)                           &(50)                       & --                                       &--                        &--\\ 
6         &03/06/02    &52339.80                    & 2.42\er0.08          & 3.23\er0.10           & 5.40\er0.38               & 7.80\er0.78          &$-0.63\pm0.04$\\
           &EW367        &(3.73)                        &(50)                           &(80)                    &(160)                         &(170)                        &--\\
7         &03/11/02     &52344.84                   & 2.30\er0.13          & 3.50\er0.10          &  --                             & --                           &$-$0.71\er0.11\\
           &EW367     &(1.59)                           &(62)                           &(80)                            &--                              &--                        &--\\
8         &04/08/02    &52372.56                    & 2.10\er0.06 (T)           & 3.39\er0.06(T)         &  --                             & --                           &$-$0.81\er0.06\\
           &6A              &(3.94)                         & 1.61\er0.16 (G1)           & 2.94\er0.13(G1)         &  --                             & --                           &--\\        
           &--                & --                               & 049\er0.16 (G2)            & 0.45\er0.13(G2) & --                             & --                           &--\\
           &--                &--                                &(50)                           &(45)                & --                                & --                             &--\\ 
9         &04/09/02    &52373.71                   & 1.90\er0.10 (T)           & --                             & --                             & --                           &--\\
           &6A              & (3.13)                        & 1.61\er0.08 (G1)           & --                              & --                             & --                           &--\\
           &--                  &--                              & $>$0.29\er0.08 (G2)     & --                              & --                             & --                           &--\\
           &--                  &--                              &(60)                              & --                              & --                             & --                           &--\\
8$+$9  &04/08+09/02       &--                      & 2.17\er0.04 (T)           & --                              & --                             & --                           &--\\           
           &--                  &--                              & 1.45\er0.13 ( G1)           & --                              & --                             & --                           &--\\
           &--                  &--                              & 0.53\er0.13 (G2)           & --                              & --                             & --                           &--\\
           &--                  &--                             & 0.19\er0.07 (G3)           & --                              & --                             & --                           &--\\
           &--                  &--                             &(36)                                    & --                              & --                             & --                           &--\\
10       &05/02/02    &52396.58                   & 2.14\er0.05 (T)        & 3.42\er0.05 (T)     & --                             & --                           &$-$0.80\er0.05\\
           &6A              & (8.36)                       & 1.64\er0.10 ( G1)           & 2.98\er0.07 (G1)          & --                             & --                           &--\\
           &--                  & --                            &  0.31\er0.09 (G2)          & 0.44\er0.07 (G2)           & --                             & --                           &--\\\
           & --                 & --                            & 0.19\er0.07 (G3)           & (35)                           & --                             & --                           &--\\
           &--                  &--                             &(40)                              & --                              & --                             & --                           &--\\
&&&&&&&\\
\hline
\end{tabular}
\end{minipage}
\end{table*}

\begin{table*}
 \centering
 \begin{minipage}{166mm}
  \contcaption{Western Jet -- ATCA Observations}
 \begin{tabular}{@{}lccccccc@{}}
 \hline\hline
Obs     & Date \&           & MJD                     & Flux$_{\rm 8.6GHz}$            & Flux$_{\rm 4.8GHz}$          & Flux$_{\rm 2.5GHz}$            & Flux$_{\rm 1.4GHz}$            & $\alpha_{\rm r}$   \\
           & array config.   & (\& obs. length)     & (\& rms)                                & (\& rms)                               & (\& rms)                                 & (\& rms)                                 & \\
           &                        &(days, hours)         &(mJy, $\mu$Jy bm$^{-1}$)    &(mJy, $\mu$Jy bm$^{-1}$)    &(mJy, $\mu$Jy bm$^{-1}$)     &(mJy, $\mu$Jy bm$^{-1}$)     &    \\  
(1)      &(2)                    &(3)                         &(4)                                          &(5)                                         &(6)                                          &(7)                                          &(8)\\           
\hline
\\
11       &05/22/02    &52416.55                   & 1.87\er0.05 (T)           & 3.27\er0.05 (T)           & --                             & --                           &$-$0.95\er0.05\\
           &6A              & (7.81)                       & 1.45\er0.07 (G1)           & 3.04\er0.08 (G1)           & --                             & --                           &--\\
           & --                 & --                            & 0.32\er0.07 (G2)           & 0.26\er0.06 (G2)           & --                             & --                           &--\\
            &--                  &--                            &(37)                               & (37)                              & --                             & --                           &--\\
12       &06/09/02    &52434.30                   & 1.65\er0.10 (T)             & 2.59\er0.15           & --                             & --                           &$-$0.77\er0.14\\
           &EW352      & (2.19)                        & 1.23\er0.11 (G1)           & (100)                           & --                             & --                           &--\\
           & --                 &  --                           & 0.42\er0.11 (G2)           &--                             & --                             & --                           &--\\
           &--                  &--                             &(100)                              & --                              & --                             & --                           &--\\
13       &07/05/02    &52460.48                   & 1.30\er0.07           & 2.51\er0.08           & --                             & --                           &$-$1.12\er0.11\\
            &1.5G                  &--                             &(47)                           &(44)                                     & --                             & --                           &--\\
14       &07/28/02    &52483.83                            & 1.18\er0.07                & 2.17\er0.09                  & --                             & --                           &$-$1.04\er0.12\\
           &1.5 G         &(5.22)                                  &(38)                             &(40)                                   & --                             & --                           &--\\
15       &08/02/02    &52488.41                            & 1.14\er0.09                & 1.95\er0.14                  & --                             & --                           &$-$0.91\er0.15\\
           &750B         &(4.38)                                  &(50)                                 &(50)                              & --                             & --                           &--\\
16       &08/30/02   &52516.16                             & 0.64\er0.07                &1.22\er0.07                   & --                             & --                           &$-$1.09\er0.15\\
           &6C            & (8.15)                                  &(38)                             &(35)                               & --                             & --                           &--\\
17       &09/17/02   &52534.20                             &1.41\er0.18 (T)           &1.67\er0.14 (T)              &2.47\er0.30              &2.99\er0.32           &$-0.43\pm0.07$\\
           &6G            & (4.67)                                  &0.90\er0.13 (G1)        &1.38\er0.10 (G1)            & (170)                       & (200)                           &--\\
           &  --               &  --                                      &0.51\er0.13 (G2)        &0.29\er0.10 (G2)           & --                             & --                           &--\\
           &  --               &  --                                     &(67)                             &(70)                               & --                             & --                           &--\\
18       &11/02/02   &52580.17                             &0.51\er0.08                 &1.11\er0.10                    &1.75\er0.30              &3.13\er0.40             &$-0.96\pm0.08$\\
           &1.5A           & (5.50)                                &(18)                             &(60)                               & (180)                       & (210)                           &--\\
19       &11/20/02    &52598.12                             &0.50\er0.12 (T)           &0.97\er0.10 (T)             &1.43\er0.20               &2.45\er0.20             &$-0.77\pm0.08$\\
           &6A              &(5.72)                                  &0.23\er0.09 (G1)        &0.71\er0.07 (G1)           & (130)                       & (150)                      &--\\
            &--                 &--                                       &0.27\er0.09 (G2)        &0.26\er0.07 ( G2)          & --                             & --                           &--\\
            &--                 &--                                       &(55)                            &(35)                               & --                             & --                           &--\\
20       &12/16/02    &52624.96 (5.47)                   &0.30\er0.07               &1.00\er0.08 (T)              &1.21\er0.19              &1.83\er0.20            &$-0.59\pm0.09$\\
           &6A              &(5.47)                                   &(50)                           &0.71\er0.06 (G1)           & (120)                             & (140)                   &--\\
           &--                  &--                                        & --                             &0.29\er0.06  (G2)           & --                             & --                           &--\\
           &--                  &--                                        & --                             &(45)                                & --                             & --                           &--\\
\\
\multicolumn{8}{c}{\bf Year 2003}\\

21       &01/26/03    &52665.83                              &$<$0.05                      &0.50\er0.07 (T)           &1.21\er0.15               &2.36\er0.40             &\\
           &6B              &(4.64)                                   &(50)                             &0.50\er0.07 (G1)            & (100)                    & (120)                           &--\\
           &--                  & --                                       & --                                &$<$0.05 (G2)                & --                             & --                           &--\\
           &--                  & --                                       & --                                &(48)                                & --                             & --                           &--\\
22       &01/27/03     &52666.83                             &0.66\er0.13 (T)           &0.78\er0.11 (T)             &1.29\er0.21              &2.34\er0.40            &\\
           &6B              &(4.66)                                   &0.33\er0.09 (G1)         &0.56\er0.08 (G1)            &(100)                             & (160)                           &--\\
           &--                  & --                                       &0.33\er0.09 (G2)         &0.22\er0.08 (G2)            & --                             & --                           &--\\
           &--                 &--                                         &(44)                                 &(40)                                   & --                             & --                           &--\\       
21$+$22 &01/26$+$27/03  &--                              &0.46\er0.08          &0.67\er0.11              &1.24\er0.16                   &2.27\er0.34           &$-0.86\pm0.19$\\    
            &--                 &--                                        &(35)                     &(35)                          &(80)                              &(150)                           &--\\     
23       &03/06/03      &52704.81                           &--                              &0.76\er0.10 (T)          &0.77\er0.13                &1.52\er0.22           &$-0.86\pm0.17$\\
           &1.5B             &(2.74)                                 &--                             &0.28\er0.17 (G1)        & (75)                             & (140)                           &--\\
           &--                   &--                                       &--                             &0.22\er0.15 (G2)         & --                             & --                           &--\\
           &--                   &--                                       &--                             &0.16\er0.10 (G3)         & --                             & --                           &--\\
           &--                   &--                                       &--                             &0.13\er0.08 (G4)         & --                             & --                           &--\\
           &--                   &--                                       &--                             &(37)                             & --                             & --                           &--\\
24       &07/25/03        &52845.49                          &--                            &0.36\er0.06 (T)              &0.56\er0.11               &1.04\er0.26           &$-0.80\pm0.23$\\
           &6D                  & (3.48)                              &--                            &0.17\er0.06 (G1)           & (100)                             & (150)                           &--\\
           &--                   &--                                        &--                            &0.19\er0.06 (G2)           & --                             & --                           &--\\    
           &--                   &--                                        &--                            &(43)                              & --                             & --                           &--\\                                                                                                                                                        
 \hline
\end{tabular} 
 Columns: 1- Observation label; 2- date of the observation and ATCA array configuration (EW 367-EW352, 375: maximum baseline of about 375 m; 750A-D: maximum baseline of about 750 m; 1.5A to 1.5G: maximum baseline of about 1500 m; 6A to 6G: maximum baseline of about 6000 m); 3- MJD of the observation, the length of the observation in hours is indicated between brackets; 4,5,6,7- flux density and errors at  at 8.6 GHz, 4.8 GHz, 2.5 GHz and 1.4 GHz. The rms noise for each frequency is reported in between brackets. Upper limits are given at 3$\times$rms; 8- radio spectral index.\\
Notes: the flux density reported in first line of each observation corresponds to the total flux of the western jet at the given frequency. When the jet was resolved, the total flux density is labelled (T) and the best fit flux densities obtained by fitting the image with multiple elliptical Gaussians are labelled (GX) with X=1,2...
\end{minipage}
\end{table*}

\begin{table*}
\caption{Western Jet -- X-rays: centroid, projected peak and tail positions.}
\label{t2}
\begin{center}
\begin{tabular}{rrrrrrrl}
\hline
\hline
 \multicolumn{1}{c}{ObsID}    & 
 \multicolumn{1}{c}{MJD}  & 
  \multicolumn{1}{c}{$\Delta$t} &
 \multicolumn{1}{c}{centroid} & 
 \multicolumn{1}{c}{Peak shift$^a$}&
 \multicolumn{1}{c}{Tail pos.$^a$} &
  \multicolumn{1}{c}{v$_{\rm app.,xte }$}&
   \multicolumn{1}{c}{v$_{\rm app,3448 }$} \\
 \multicolumn{1}{c}{}    & 
 \multicolumn{1}{c}{(days)}  &
 \multicolumn{1}{c}{(days)}  &
 \multicolumn{1}{c}{(\arcsec)}&     
  \multicolumn{1}{c}{(\arcsec)}&       
 \multicolumn{1}{c}{(\arcsec)}  &
 \multicolumn{1}{c}{(mas days$^{-1}$) }&
  \multicolumn{1}{c}{(mas days$^{-1}$) } \\  
   \multicolumn{1}{c}{(1)}    & 
    \multicolumn{1}{c}{(2)}    & 
     \multicolumn{1}{c}{(3)}    & 
      \multicolumn{1}{c}{(4)}    & 
       \multicolumn{1}{c}{(5)}    & 
        \multicolumn{1}{c}{(6)}    & 
         \multicolumn{1}{c}{(7)}    & 
          \multicolumn{1}{c}{(8)}   \\ 
\hline
3448        & 52344.62\er0.14    &1266.81   &22.6\er0.5       &22.75\er0.5$^b$    &  19.0               &17.9\er0.4    &--\\

3672        &  52444.38\er0.10   &1366.57   &23.2\er0.5       &0.52\er0.12    &  18.75                     &17.0\er0.4       &5.2\er1.2 \\

3807        &   52541.83\er0.14  &1464.02   &23.3\er0.5       &0.7\er0.12      &  18.25             &15.9\er0.3         &3.5\er0.6        \\
                &                               &                           &                        &                      &  18$^\ast$        &                                 &\\
4368        & 52667.19\er0.12    &1589.38   &23.9\er0.5       &0.85\er0.22$^c$      &18.25             &15.1\er0.3          &2.6\er0.7 \\     
                &                               &                           &                        &                      &17.75$^\ast$    &                  &\\
5190        &  52935.30\er0.27   &1857.49  &24.0\er0.5        &0.84\er0.07$^{c,d}$      &   16.75       &12.9\er0.3         &3.0\er1.0$^e$           \\
                &                               &                           &                        &                       &   16.5$^\ast$          &                &\\
\hline
\end{tabular}
\end{center}
Columns: 1- \Cha~observation ID; 2- Modified Julian date of the observation,  the error corresponds to half the length of the observation; 3- time from the 1998 X-ray flare (MDJ=51077.8); 4- projected distance of the X-ray centroid (identified using \texttt{wavdetect}, see Sec. 3) from \xte;  5- projected separation of the X-ray peak of the western jet (in the brightness profile) with respect to the initial peak position in ObsID 3448. A KS test is used to compare the western jet position between ObsID 3448 and the other observations. In all cases, the probability of a zero offset is $<10^{-11}$;  6- projected distance of the X-ray tail from \xte. Uncertainties are at 90\% confidence bounds; 7- apparent advance velocity of the X-ray centroid with respect to \xte; 8- apparent velocity of the X-ray peak with respect to its position in first detection in ObsID 3448. \\
$^a$: the bin size is 0.25\arcsec.\\
$^b$: initial distance of the X-ray peak of western jet's brightness profile from \xte.\\
$^c$: the KS test between ObsID 3448 and the last two observations (4368 and 5190) gives a low probability that the two samples are drawn from the same distribution.\\
$^d$: the peak offset measured by the KS differs from the actual offset (1.75\arcsec\er0.35\arcsec) between the  peaks.\\
 $^e$: calculated from the actual offset  between the peaks (1.75\arcsec\er0.35\arcsec).\\
$^\ast$: position of the last 1 count bin (located in a train of bins with more than 1 count).
\end{table*}

\begin{table*}
\caption{Western Jet -- X-ray Spectral Analysis: Best Fit Model}
\label{t3spec}
\begin{center}
\begin{tabular}{ r c c r r r r r}
\hline
\hline
  \multicolumn{1}{c}{ObsID}&
  \multicolumn{1}{c}{Exp. time}&
  \multicolumn{1}{c}{Counts}&
 \multicolumn{1}{c}{$\Gamma$}&
 \multicolumn{1}{c}{norm$_{\rm \Gamma,Tot}$}&
 \multicolumn{1}{c}{F$_{\rm 0.3-8keV,Tot}^{a}$}&
  \multicolumn{1}{c}{$\Gamma_{\rm Tail}$}&
  \multicolumn{1}{c}{F$_{\rm 0.3-8keV,tail}^{a}$}\\
     \multicolumn{1}{c}{(1)}    & 
    \multicolumn{1}{c}{(2)}    & 
     \multicolumn{1}{c}{(3)}    & 
      \multicolumn{1}{c}{(4)}    & 
       \multicolumn{1}{c}{(5)}   &
      \multicolumn{1}{c}{(6)}  &
       \multicolumn{1}{c}{(7)} &
       \multicolumn{1}{c}{(8)}\\
\hline
  3448 & 24.39  & 414  & 1.85$^{+0.11}_{-0.10}$   &6.10$^{+0.64}_{-0.61}$   &3.46$^{+0.17}_{-0.22}$   &2.04$^{+0.18}_{-0.40}$  &0.8\\
  3672 & 17.66  & 238  & 1.79$^{+0.13}_{-0.14}$   &4.94$^{+0.64}_{-0.60}$   &2.90$^{+0.19}_{-0.24}$   &1.61$^{+0.30}_{-0.30}$  &0.5\\
  3807 & 24.44  & 197  & 2.15$^{+0.16}_{-0.14}$   &4.14$^{+0.57}_{-0.52}$   &2.05$^{+0.17}_{-0.16}$   &2.10$^{+0.35}_{-0.35}$  &0.4\\
  4368 & 22.40  & 110  & 1.98$^{+0.22}_{-0.21}$   &2.06$^{+0.44}_{-0.36}$   &1.12$^{+0.11}_{-0.13}$   &2.09$^{+0.40}_{-0.40}$   &0.4\\
  5190 & 46.55  & 145  & 1.93$^{+0.18}_{-0.18}$   &1.28$^{+0.23}_{-0.21}$   &0.62$^{+0.04}_{-0.06}$   &1.67$^{+0.25}_{-0.25}$  &0.36\\
\hline\end{tabular}
\end{center}
Columns: 1- \Cha~observation ID; 2- exposure time of the observation in ksec after filtering for background flares; 3- number of photons in the extraction region in the 0.3-8.0 keV band; 4- 0.3-8.0 keV best-fit photon index of the whole jet emission; 5- power-law normalization of the whole jet emission in units of 10$^{-5}$ photons cm$^{-2}$ s$^{-1}$; 6- unabsorbed 0.3-8.0 keV flux in whole jet region in units of 10$^{-13}$ ergs cm$^{-2}$ s$^{-1}$; 7- 0.3-8.0 keV best-fit photon index of the emission in the tail region; 8- unabsorbed 0.3-8.0 keV flux in tail region in units of 10$^{-13}$ ergs cm$^{-2}$ s$^{-1}$. \\
$^a$:  a photoelectric model (\texttt{xswabs} in Sherpa) with the absorbing column fixed to the Galactic value, N$_{\rm H}=9\times 10^{21}$ cm$^{-2}$, is included in the model. 
\end{table*}

\begin{table*}
\caption{Western Jet -- ATCA observations: angular separation \&  apparent velocity.}
\label{t4radsep}
\begin{center}
\begin{tabular}{l r r l l}
\hline
\hline
  \multicolumn{1}{c}{Obs} &
  \multicolumn{1}{c}{MJD} &
  \multicolumn{1}{c}{Separ.} &
  \multicolumn{1}{c}{v$_{\rm app.,xte}$} &
  \multicolumn{1}{c}{v$_{\rm app,obs1/01}$} \\
    \multicolumn{1}{c}{} &
  \multicolumn{1}{c}{days} &
  \multicolumn{1}{c}{arcsec} &
   \multicolumn{1}{c}{mas day$^{-1}$} &
  \multicolumn{1}{c}{mas day$^{-1}$} \\
     \multicolumn{1}{c}{(1)}    & 
    \multicolumn{1}{c}{(2)}    & 
     \multicolumn{1}{c}{(3)}    & 
      \multicolumn{1}{c}{(4)} &   
        \multicolumn{1}{c}{(5)} \\
\hline
\multicolumn{5}{c}{\bf Year 2001}\\
1/01   &51949.99\er0.15  &20.7\er0.5 &23.7\er0.6 &--\\
\multicolumn{5}{c}{\bf Year 2002}\\
  1 & 52290.86\er0.07 & 22.55\er0.33 & 18.6\er0.3   &5.4\er1.8\\
  2 & 52292.86\er0.15 & 22.12\er0.20 & 18.2\er0.2   &4.1\er1.6\\
  3 & 52303.72\er0.12 & 22.75\er0.19 & 18.6\er0.2   &5.8\er1.5\\
  4 & 52306.90\er0.10 & 22.58\er0.19 & 18.4\er0.2   &5.3\er1.5\\
  5 & 52319.73\er0.08 & 22.04\er0.38 & 17.7\er0.3   &3.6\er1.7\\
  6 & 52339.80\er0.08 & 22.32\er0.58 & 17.7\er0.5   &4.2\er2.0\\
  8 & 52372.56\er0.08 & 23.30\er0.20 & 18.0\er0.2   &6.1\er1.3\\
  9 & 52373.71\er0.07 & 23.23\er0.25 & 17.9\er0.2   &6.0\er1.3\\
  10 & 52396.58\er0.17 & 23.21\er0.21 & 17.6\er0.2   &5.6\er1.2\\
  11 & 52416.55\er0.17 & 23.24\er0.29 & 17.4\er0.2   &5.5\er1.2\\
  12 & 52434.30\er0.05 & 22.70\er0.53 & 16.7\er0.4   &4.1\er1.5\\
  13 & 52460.48\er0.12 & 23.08\er0.31 & 16.7\er0.2   &4.6\er1.1\\
  14 & 52483.38\er0.11 & 23.21\er0.33 & 16.5\er0.2   &4.7\er1.1\\
  15 & 52488.41\er0.09 & 22.87\er0.41 & 16.2\er0.3   &4.0\er1.2\\
  16 & 52516.29\er0.17 & 23.00\er0.32 & 16.0\er0.2   &4.1\er1.0\\
  17 & 52534.20\er0.10 & 23.50\er0.37 & 16.1\er0.3   &4.8\er1.1\\
  18 & 52580.17\er0.11 & 22.74\er0.50 & 15.1\er0.3   &3.2\er1.1\\
  19$^\ast$ & 52598.12\er0.12 & 23.02\er0.38 & 15.1\er0.2   &3.7\er1.0\\
  20 & 52624.96\er0.11 & 23.14\er0.36 & 15.0\er0.2   &3.6\er0.9\\
  \multicolumn{5}{c}{\bf Year 2003}\\
  21+22 & 52666.33\er0.33 & 23.94\er0.46 & 15.1\er0.3   &4.5\er0.9\\
  23 & 52704.81\er0.06 & 23.72\er0.53 & 14.6\er0.3 &4.0\er1.0\\
  24$^\ast$ & 52845.49\er0.07 & 23.50\er0.84 & 13.3\er0.5   &3.1\er1.1\\ 
\hline
\end{tabular}
\end{center}
Columns: 1- ATCA observation; 2- Modified Julian date of the observation, the error corresponds to half the length of the observation; 3- angular separation of the brightest radio component from \xte~(see the text). Unless differently indicated, the positions have been measured using the radio maps at 8.64 GHz; 4- average apparent advance velocity with respect to \xte;  5- apparent advance velocity with respect to the position at the time of the first radio detection in February 2001 (obs1/01).\\
$^\ast$:  measured using the 4.8 GHz maps.\\
\end{table*}

\begin{table*}
\caption{X-ray eastern Jet: centroid angular separation and apparent velocity}
\label{t5}
\begin{center}
\begin{tabular}{l r r r l}
\hline
\hline
  \multicolumn{1}{c}{ObsID} &
  \multicolumn{1}{c}{MJD} &
  \multicolumn{1}{c}{centroid} &
  \multicolumn{1}{c}{v$_{\rm app.,xte}$}&
   \multicolumn{1}{c}{v$_{\rm app.,679}$}  \\
  \multicolumn{1}{c}{} &
  \multicolumn{1}{c}{days} &
  \multicolumn{1}{c}{arcsec} &
 \multicolumn{1}{c}{mas days$^{-1}$} &
  \multicolumn{1}{c}{mas days$^{-1}$} \\
       \multicolumn{1}{c}{(1)}    & 
    \multicolumn{1}{c}{(2)}    & 
     \multicolumn{1}{c}{(3)}    &
          \multicolumn{1}{c}{(4)}  &  
      \multicolumn{1}{c}{(5)}     \\
\hline
   679  &51704.54\er0.03  &21.3\er0.5         &34.0\er0.7   &--\\
  1845 & 51777.41\er0.03 & 22.7\er0.5        & 32.4\er0.7  &19\er10\\
  1846 & 51798.25\er0.03 & 24.2\er0.5        & 33.6\er0.7  &31\er7\\
  3448 & 52344.63\er0.14 & 28.6\er0.5        & 22.6\er0.4  &11\er1\\
  3807$^a$ & 52541.83\er0.14 & 28.6\er1.0 & 19.5\er0.7  &8.7\er0.8\\
  5190$^a$ & 52935.30\er0.27 & 33.3\er1.5 & 17.9\er0.8  &10\er1.0\\
\hline\end{tabular}
\end{center}
Columns: 1- \Cha~observationID; 2- Modified Julian date of the observation, the error corresponds to half the length of the observation; 3- projected distance of the X-ray centroid (identified using \texttt{wavdetect}) from \xte; 4- apparent advance velocity of the X-ray centroid with respect to \xte;  5- apparent advance velocity of the X-ray centroid with respect to the position in the first detection (ObsID 679).\\
$^a$: the centroid position was measured using the ds9 dax. 
\end{table*}
\clearpage

\begin{figure*}
\centering
\includegraphics[scale=0.4]{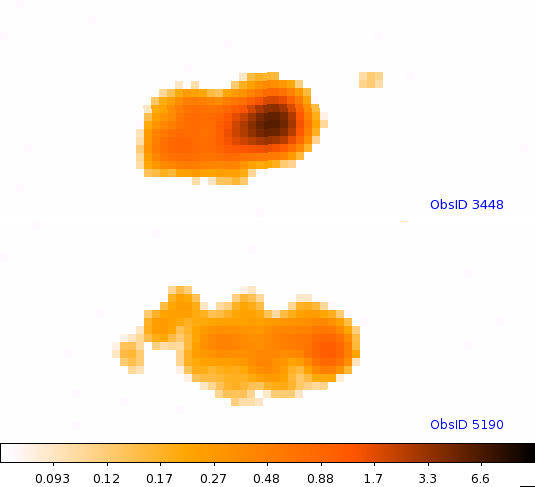}
\includegraphics[scale=0.4]{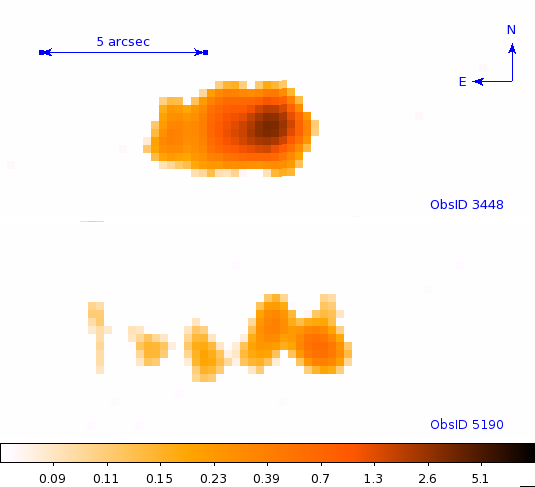}
\caption{Smoothed \Cha~ACIS-S images of the western jet of XTE J1550$-$564 in the 0.3-2.5 keV energy band (left panel) and in the 2.5-8.0 keV energy band (right panel) in the first (upper panels) and last (lower panels) observations. The pixel size is set to half of the original ACIS pixel and is equal to 0.246\arcsec. The color scale, in count units, is logarithmic.}
\label{f2}
\end{figure*}

\begin{figure*}
\centering
\includegraphics[scale=0.75]{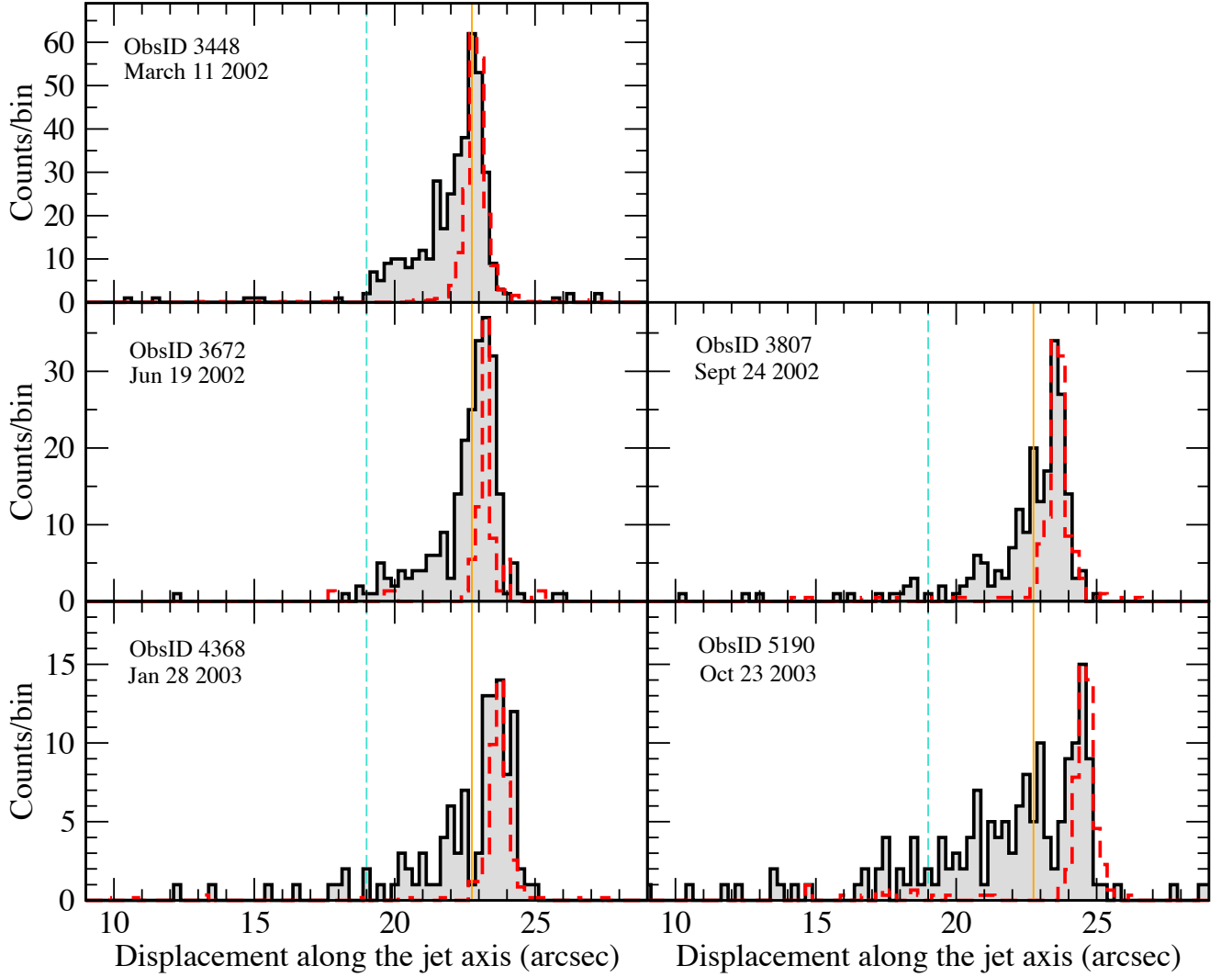}
\caption{Longitudinal profiles of the western X-ray jet in the 0.3-8 keV energy range for the five \Cha detections, using a bin size of 0.246\arcsec. The dashed red line is the profile of \xte~at the same epoch, which has been shifted and re-normalized to match the western X-ray jet peak. The vertical solid orange line and dashed cyan line mark the positions of the peak and of tail, respectively, at the epoch of the first detection.}
\label{f4}
\end{figure*}

\begin{figure*}
\centering
\includegraphics[scale=0.5]{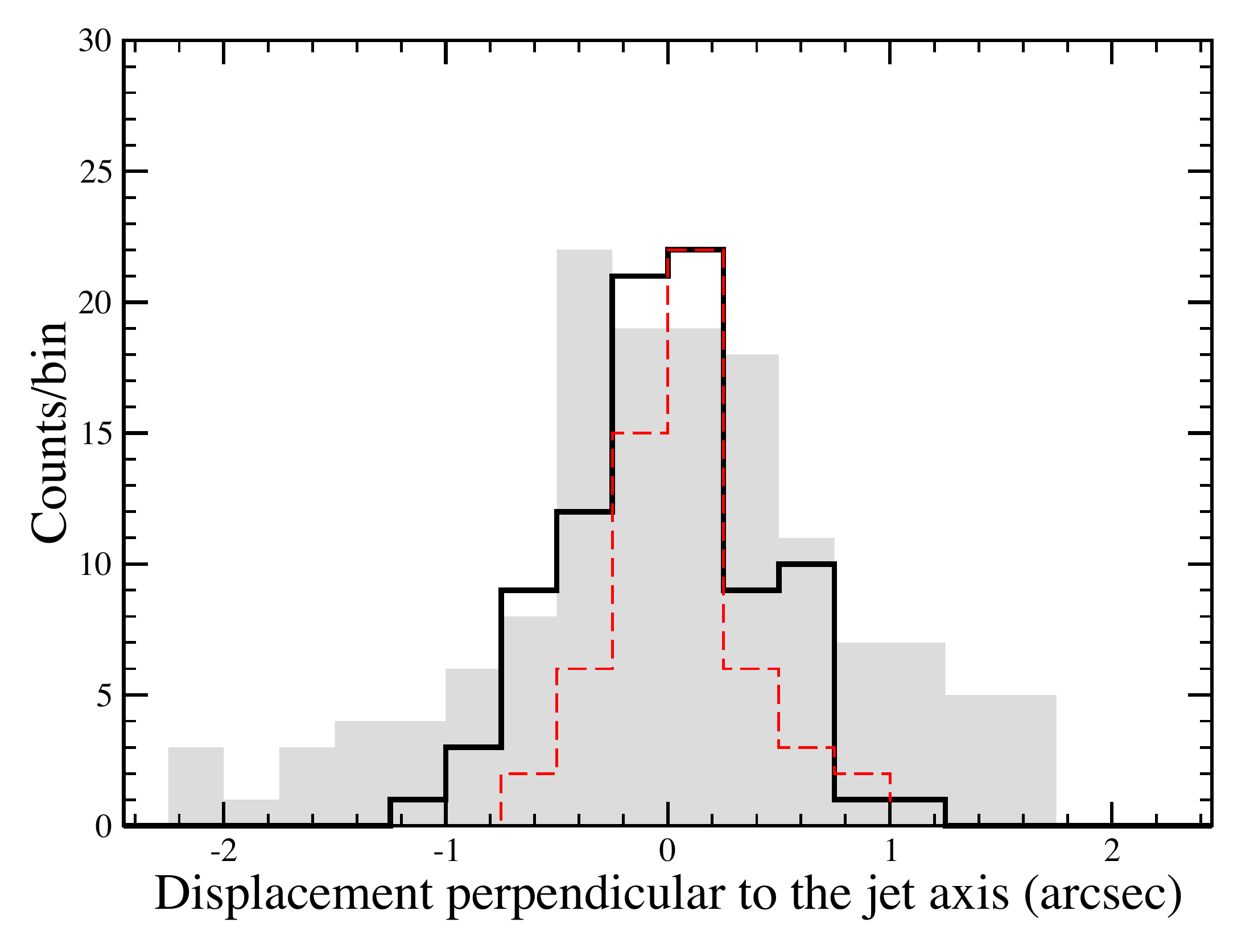}
\caption{ Profiles of the western jet perpendicular to its main axis in the 0.3-8 keV energy range: the shaded grey area is for the last observation in October 2003 (ObsID 5190), the solid black line is for the \Cha~observation in March 2002 (ObsID 3448) and  the dashed red line is the profile of \xte~(in the last observation). The latter two have been rescaled to match the peak of the emission of the western jet in October 2003. The bin size is 0.246\arcsec.}
\label{f4a}
\end{figure*}

\begin{figure*}
\centering
\includegraphics[scale=0.42]{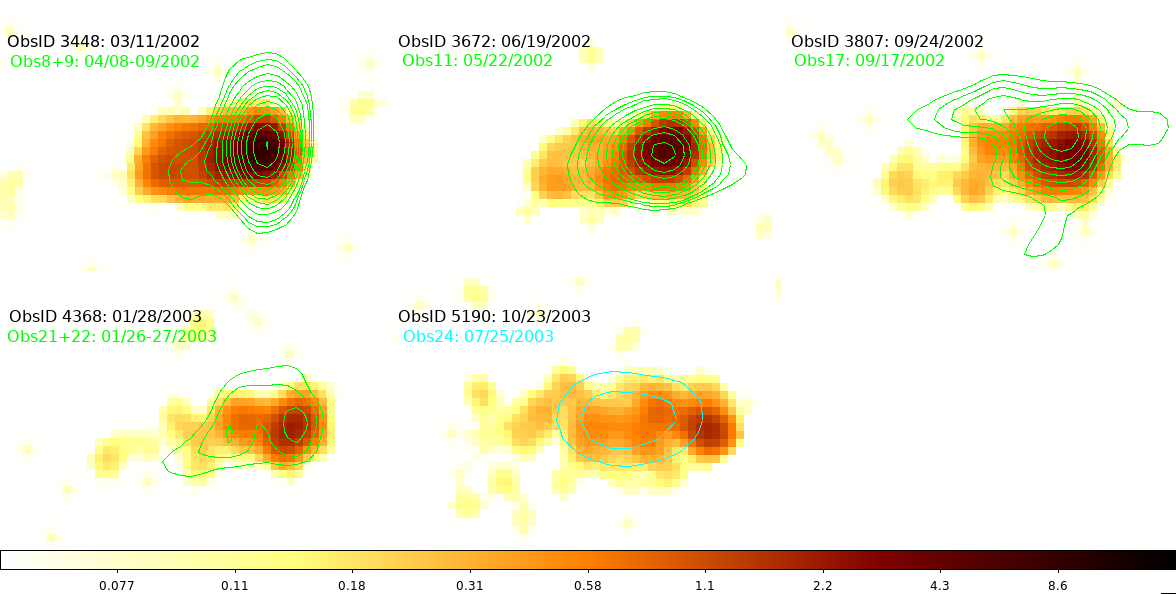}
\caption{Comparison between the X-ray and radio morphologies of the western jet. The X-ray images are for the 0.3-8 keV band.  A common logarithmic scale (in count units) is set for the five observations. The overlaid radio contours are for the closest-in-time ATCA observation at 8.6 GHz (in green) or 4.8 GHz (in cyan). The radio contour levels are 3, 4, 5, 7, 9, 11, 13, 15, 18, 20, 25, 30, 35, 40 the rms noise level.}
\label{f9}
\end{figure*}

\begin{figure*}
\centering
\includegraphics[scale=0.65]{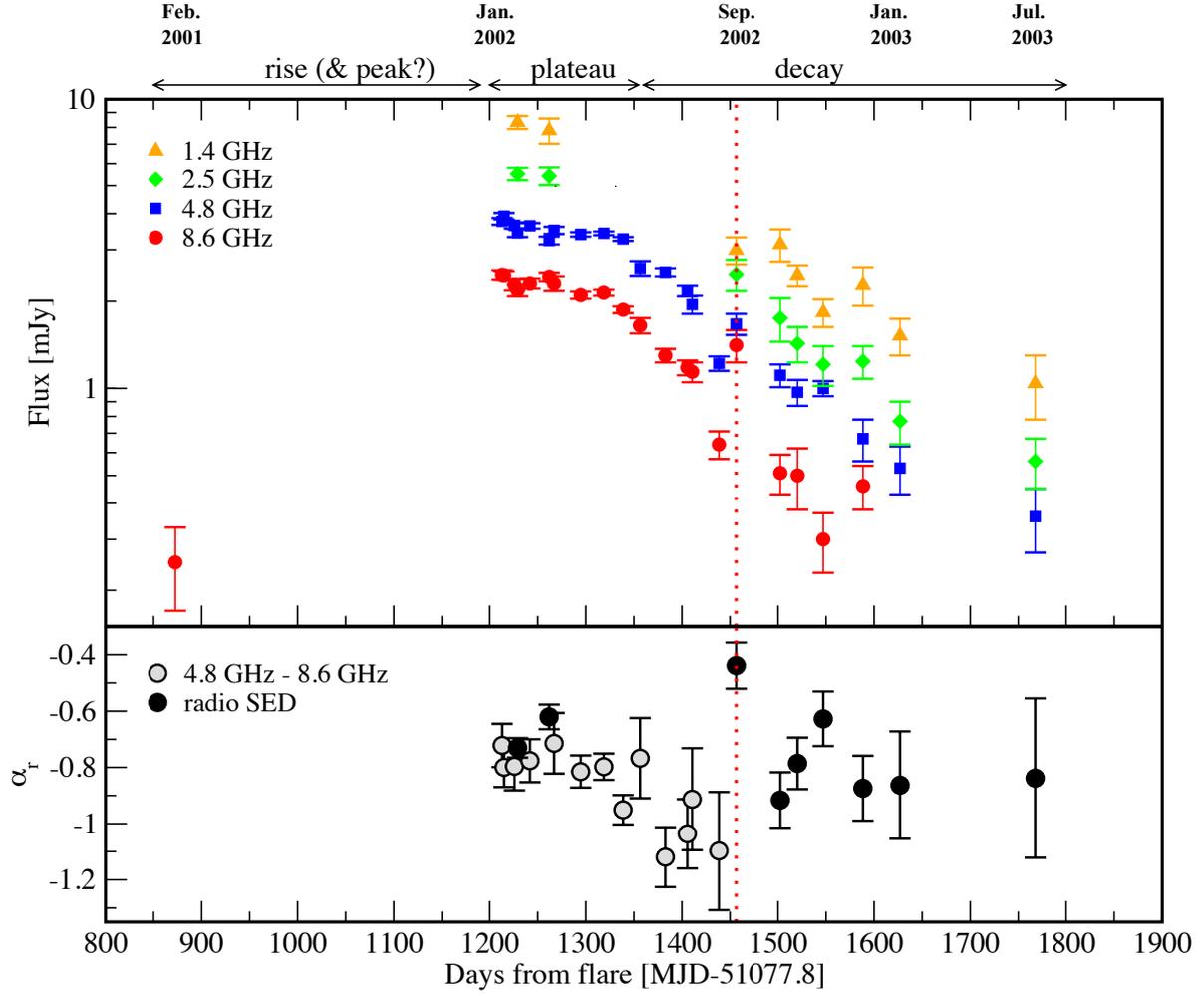}
\caption{Upper panel: radio lightcurves at 8.6 GHz (red points), 4.8 GHz (blue squares), 2.5 GHz (green diamonds) and 1.3 GHz (yellow triangles) of the western jet of \xte. The dotted vertical line marks the observed re-flare at 8.6 GHz. Lower panel: radio spectral indexes, $\alpha_r$: the black solid dots are the $\alpha_r$ obtained by fitting of the radio (3 or 4 frequencies) SED, the empty dots are derived from the 4.8 to 8.6 GHz spectrum.}
\label{f10a}
\end{figure*}

\begin{figure*}
\centerline{\includegraphics[scale=0.65]{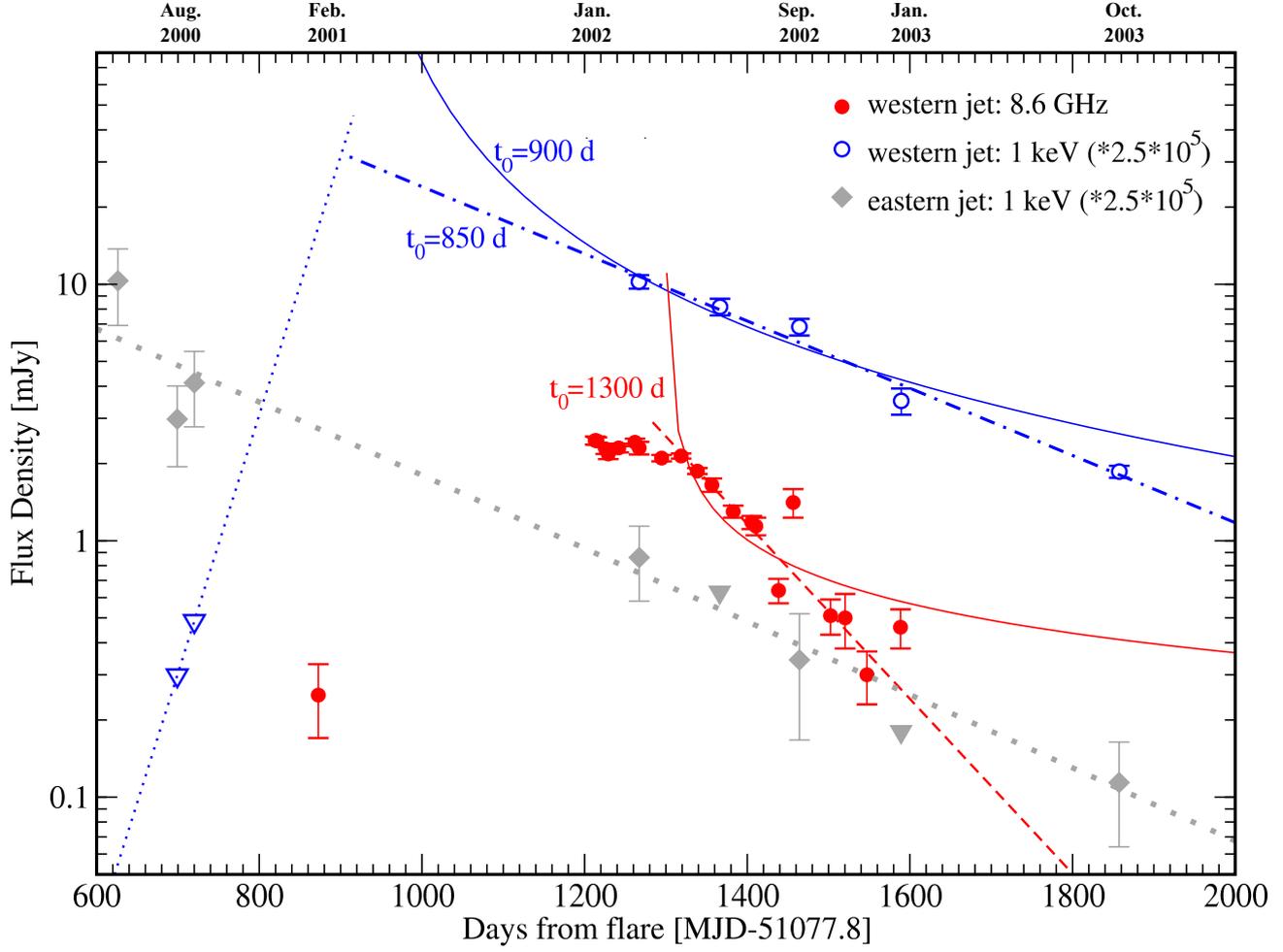}}
\caption{X-ray (1 keV, blue empty circles) and radio (8.6 GHz, red solid circles) lightcurves of the western jet and X-ray lightcurve (1 keV, grey solid diamonds) of the eastern jet. The X-ray upper limits (90\% confidence level) of the western (blue empty triangles) and eastern (grey solid triangles) jets are also shown (see text). For the purpose of comparison with the radio fluxes, the X-ray fluxes have been multiplied  by a factor of 2.5$\times$10$^{5}$. The blue dotted-dashed line, the red dashed line and the grey dotted line are the exponential decay fits of each dataset, while the solid lines correspond to the power-law fits. The dotted blue line marks  the rising time of the X-ray flux of the western jet set by the upper limits.}
\label{f12}
\end{figure*}

\begin{figure*}
\centering
\includegraphics[scale=0.7]{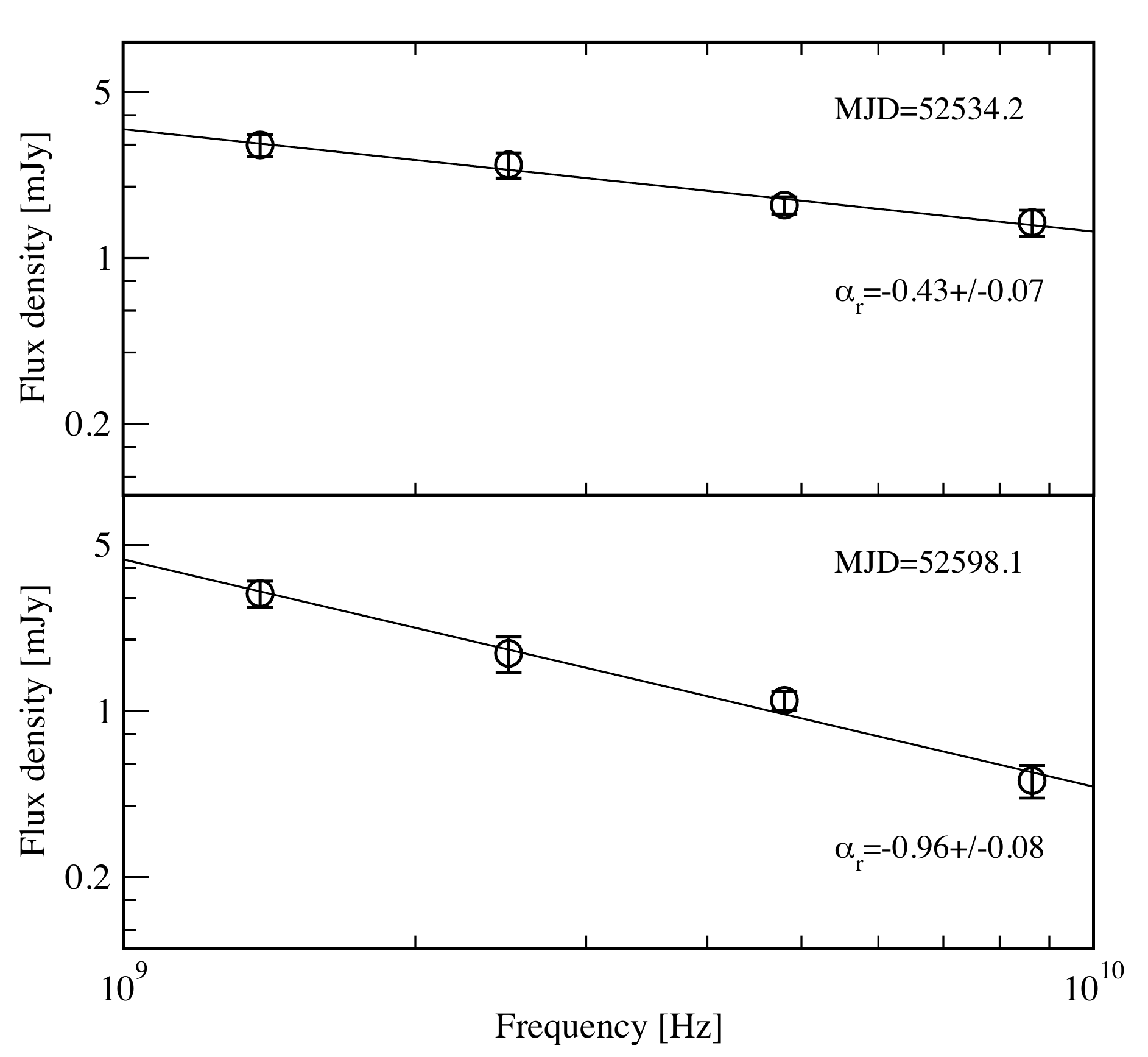}
\caption{Fit of the radio spectrum of obs17 (upper panel) and obs18 (lower panel). The flux densities are reported in Table \ref{t1}.}
\label{f11a}
\end{figure*}

\clearpage

\begin{figure*}
\centering
\includegraphics[scale=0.3]{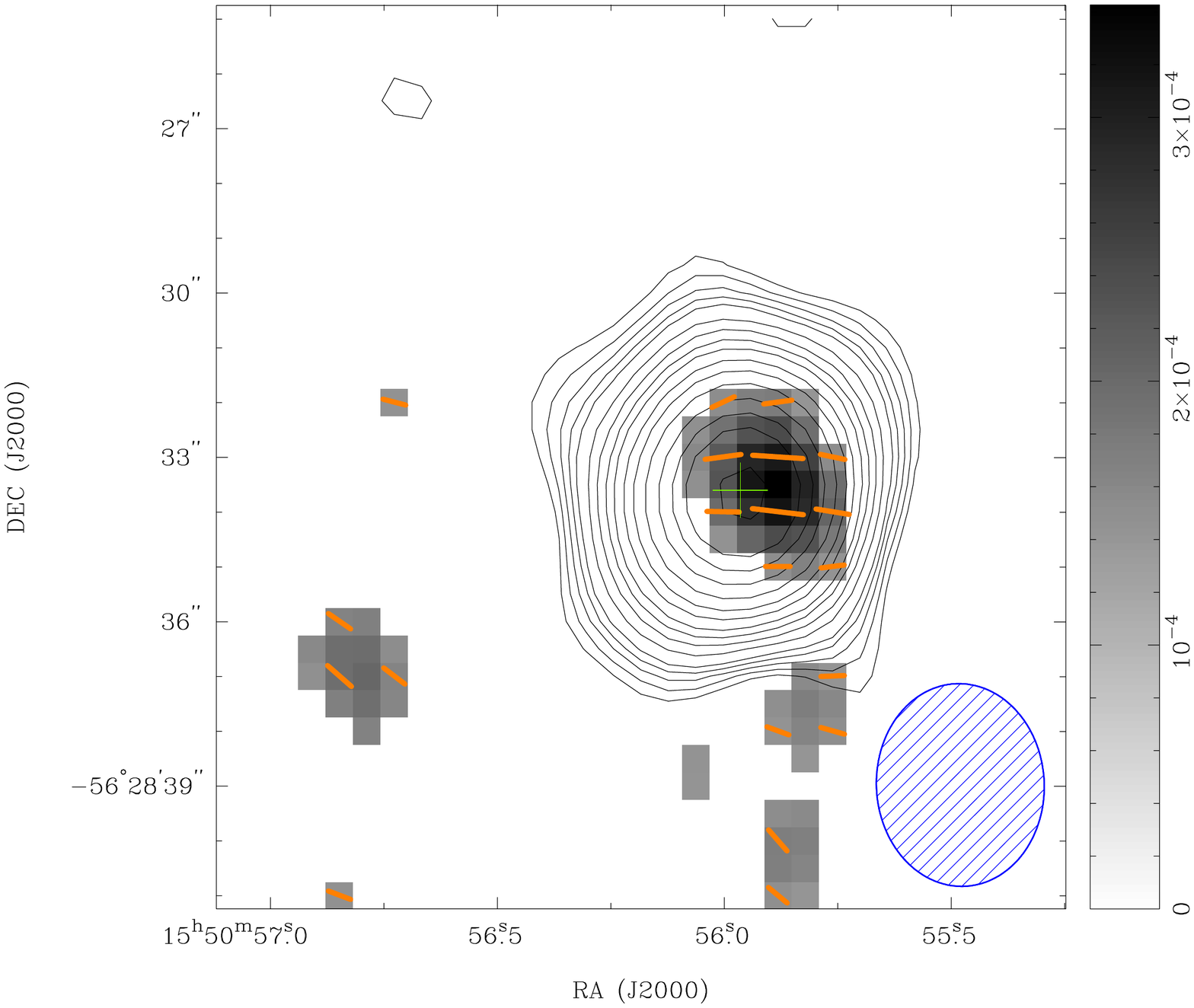}
\includegraphics[scale=0.3]{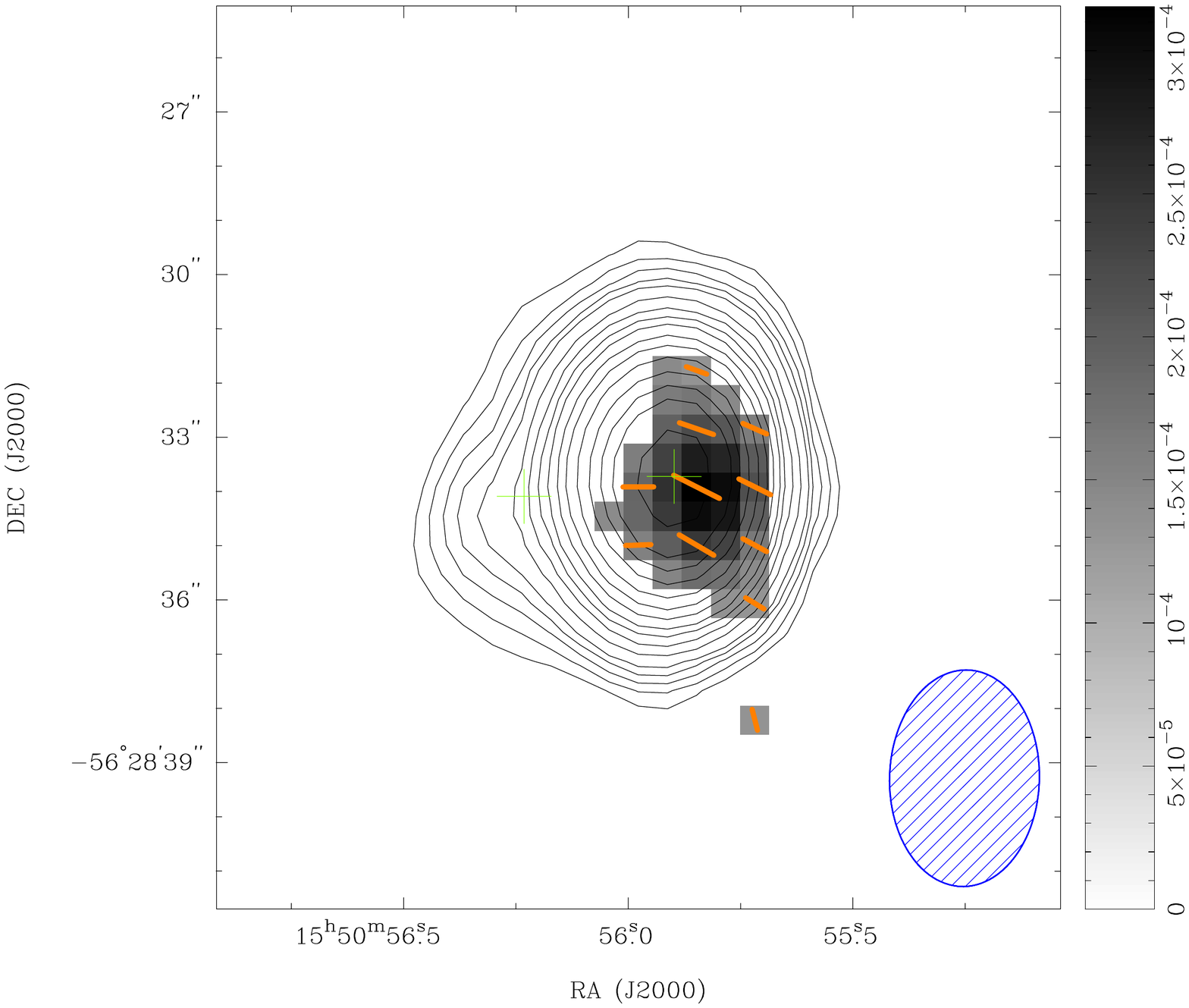}
\includegraphics[scale=0.3]{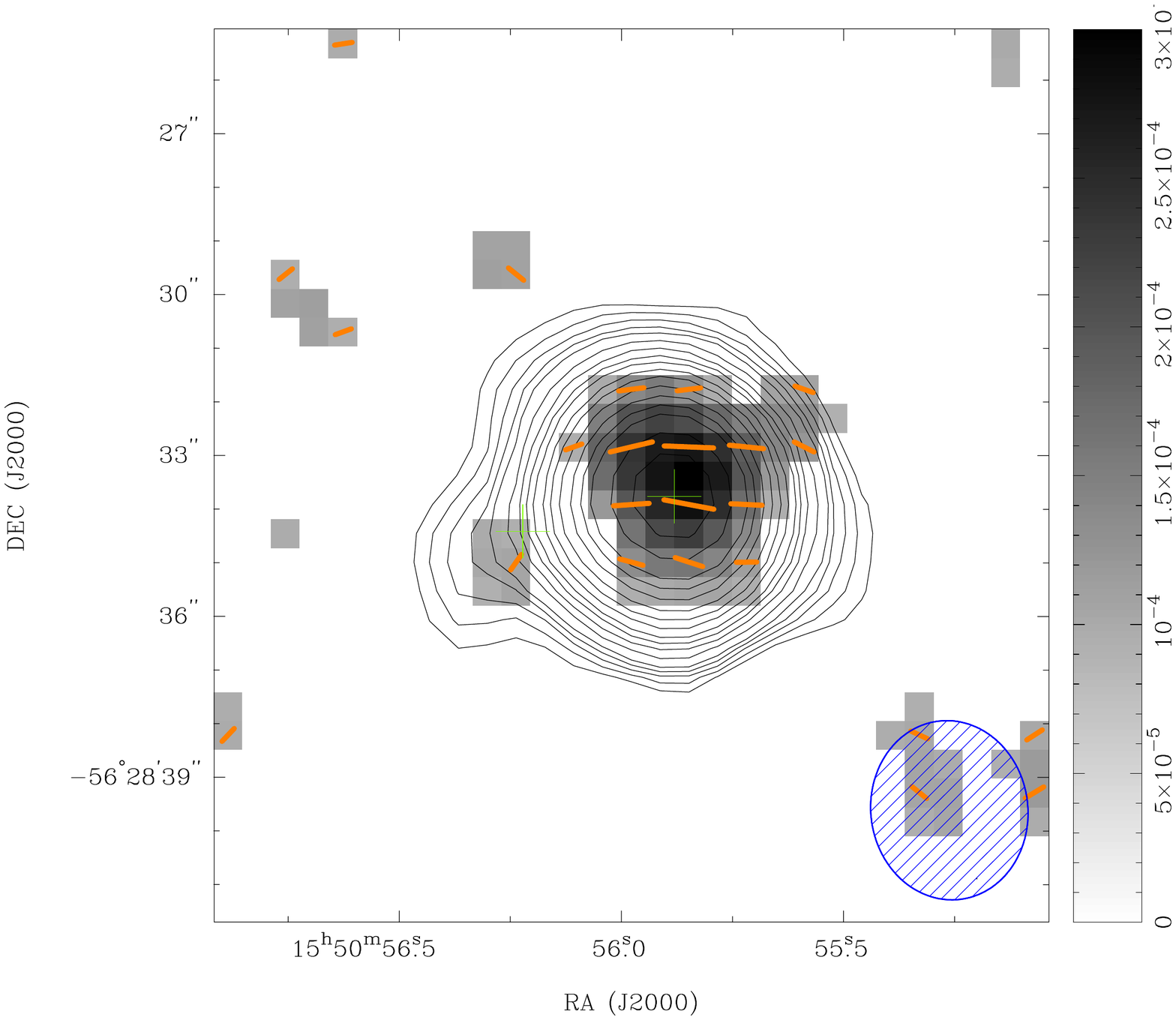}
\includegraphics[scale=0.3]{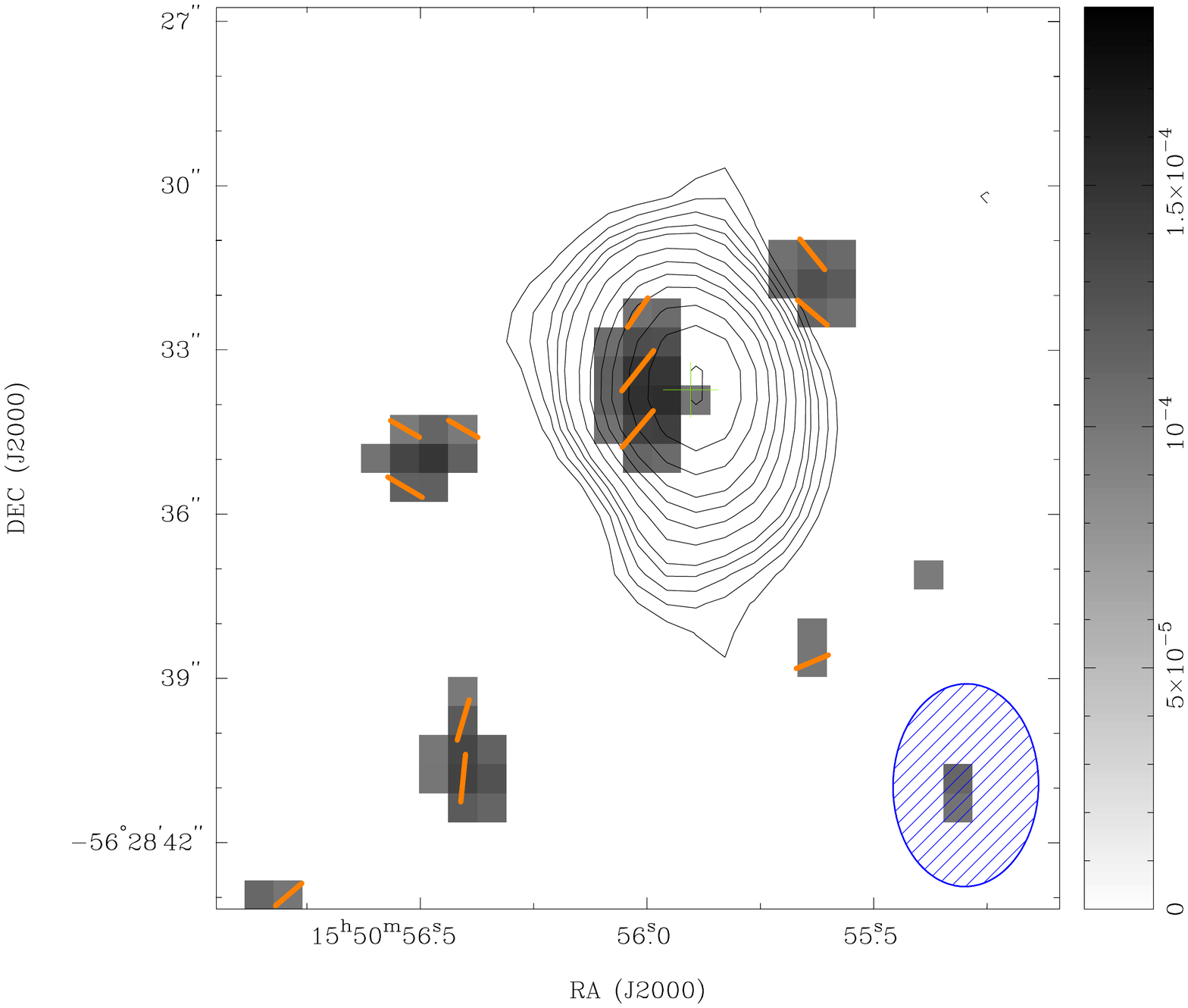}
\caption{The 4.8 GHz polarization map of the western jet of ATCA observations 4 (upper left panel), 10 (upper right panel), 11 (lower left panel) and 16 (lower right panel) performed in 2002 (see in Table \ref{t1}). The linear polarization (grey scale) is shown only in regions where it has $\geq$3 $\sigma$ significance, with the exception of observation 10 where a $\geq$4 $\sigma$ threshold has been used. The total intensity contours are overlaid. The orange lines correspond to the orientation of the EVPA. No correction for the Faraday rotation within our Galaxy has been applied.}
\label{f11b}
\end{figure*}

\begin{figure*}
\centering
\includegraphics[scale=0.31]{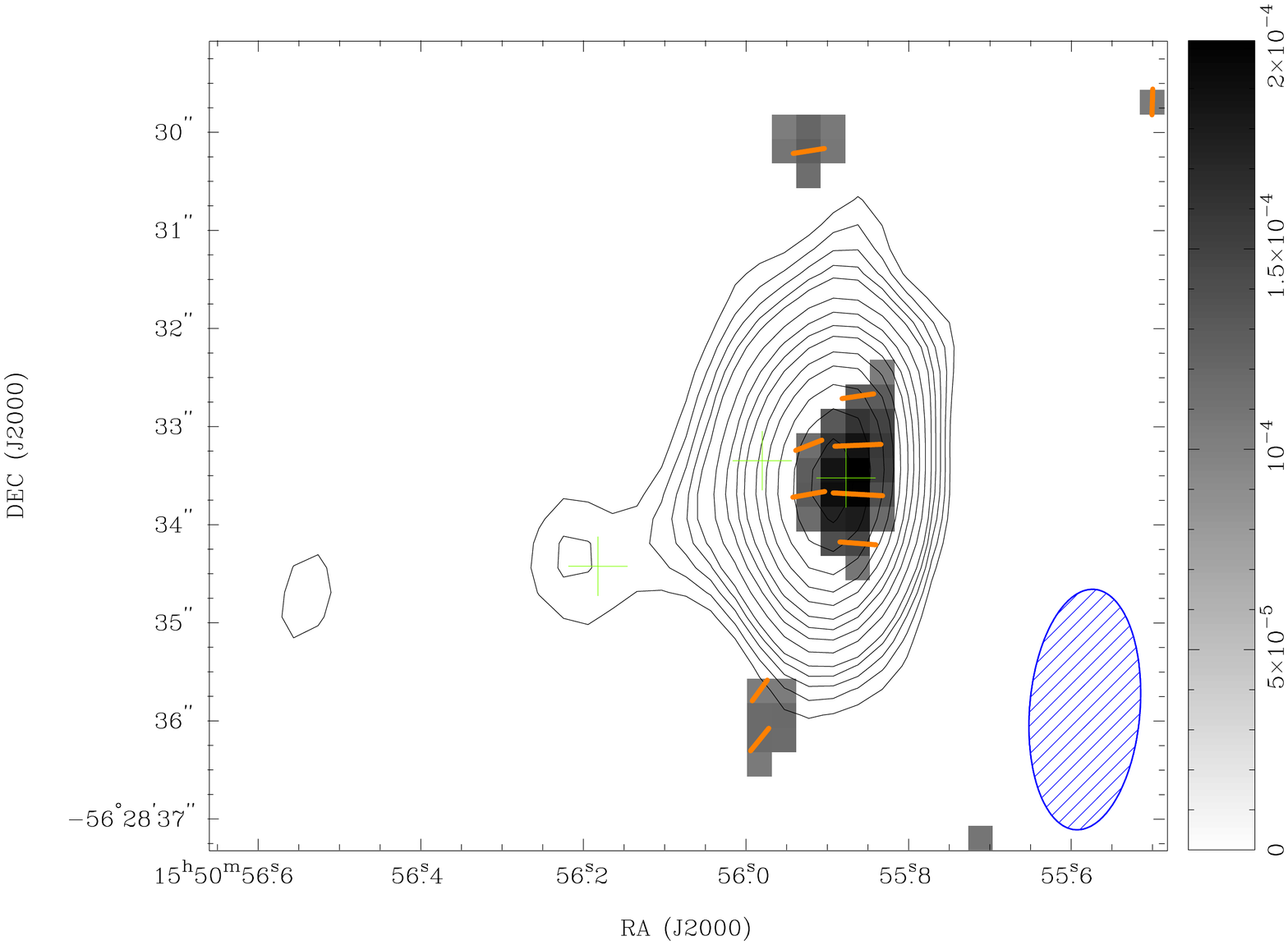}
\includegraphics[scale=0.31]{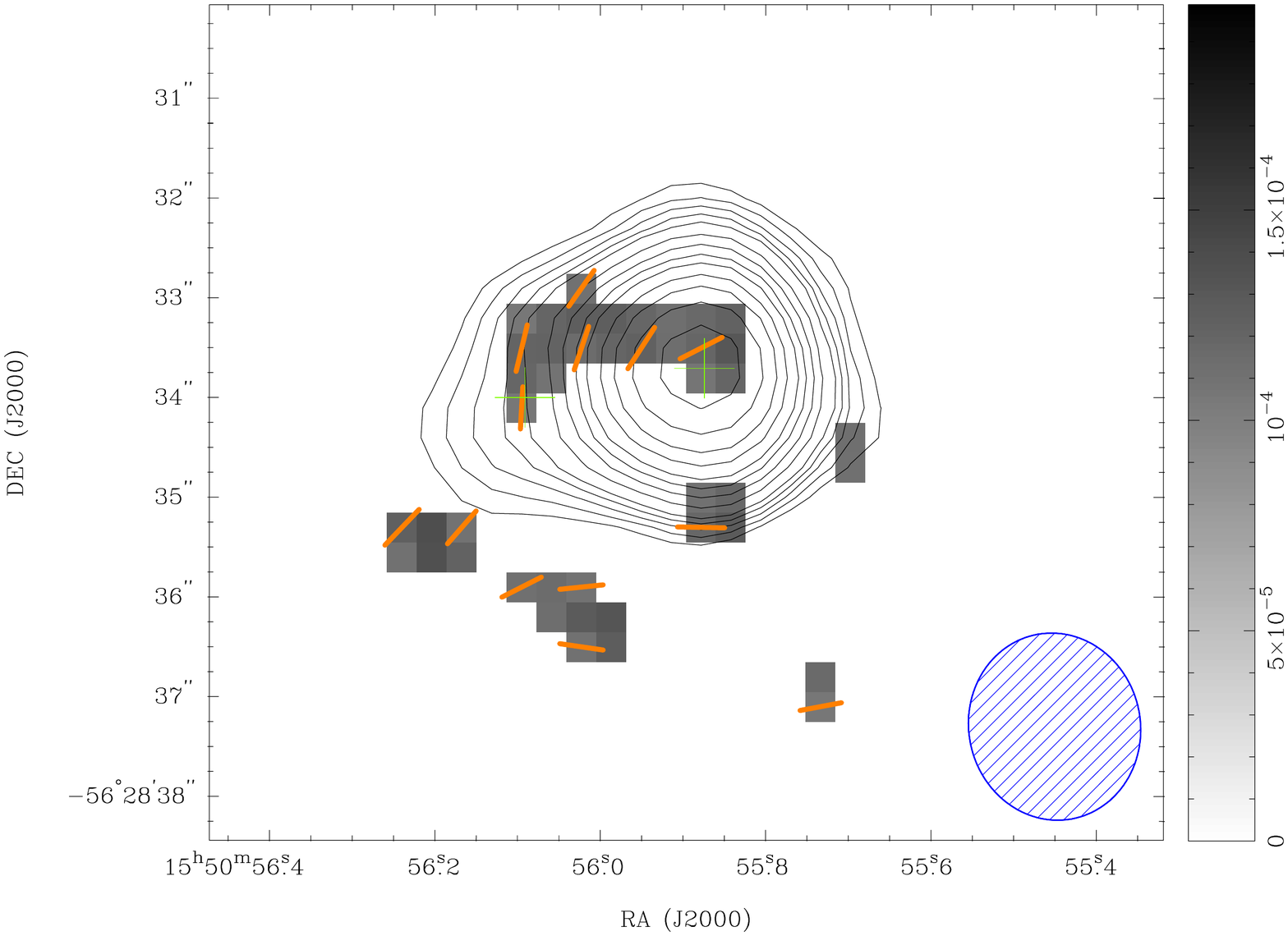}
\caption{The 8.6 GHz polarization map of the western jet of observations 8 and 9 merged (left panel, obs8+9 in Table \ref{t1}) and of the observation 11 (right panel). The linear polarization (grey scale) is shown only in regions where it has $\geq$3 $\sigma$ significance. The total intensity contours are overlaid. The orange lines correspond to the orientation of the EVPA. No correction for the Faraday rotation within our Galaxy has been applied.}
\label{f11c}
\end{figure*}

\begin{figure*}
\centering
\includegraphics[scale=0.5]{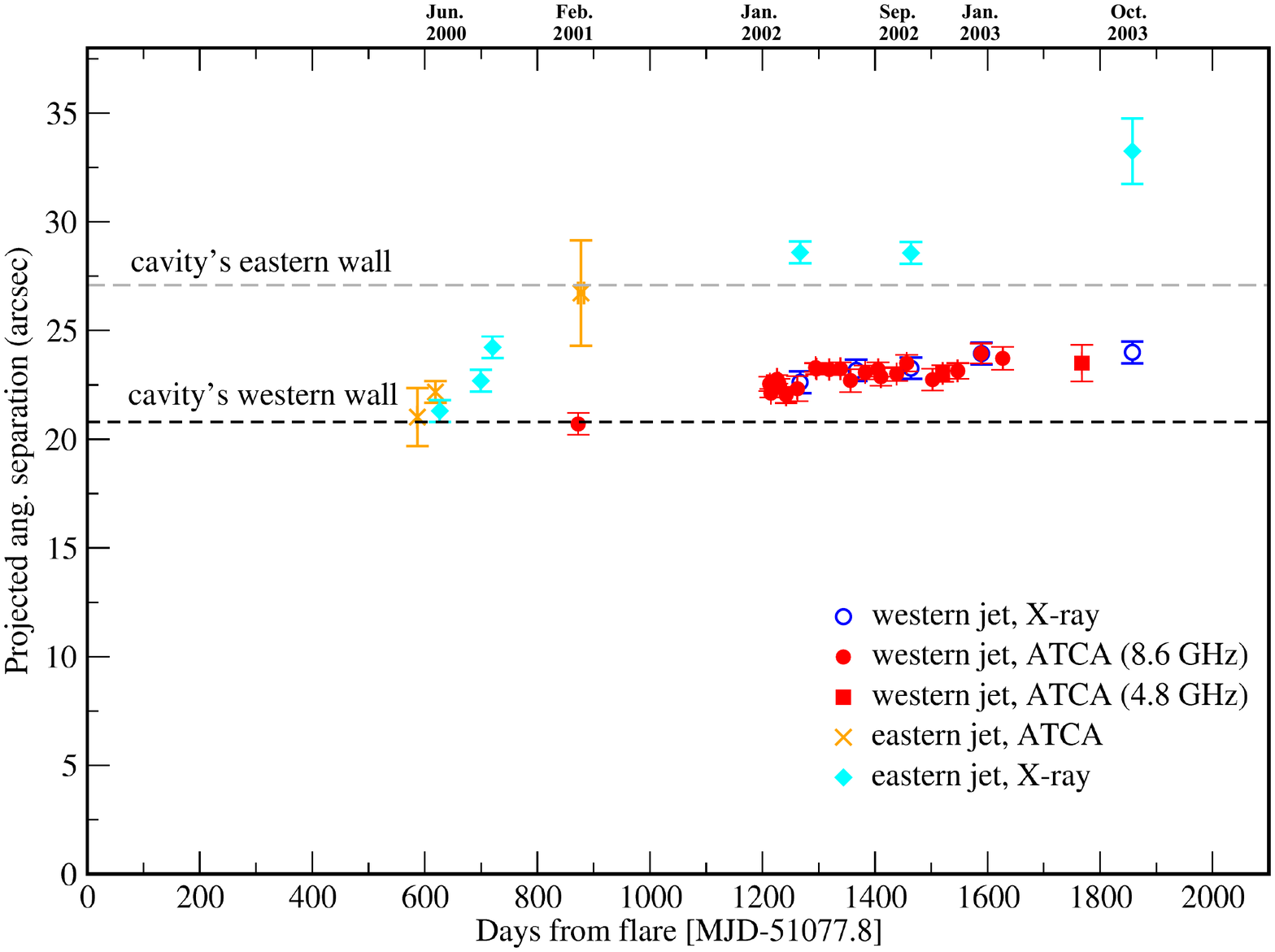}
\includegraphics[scale=0.5]{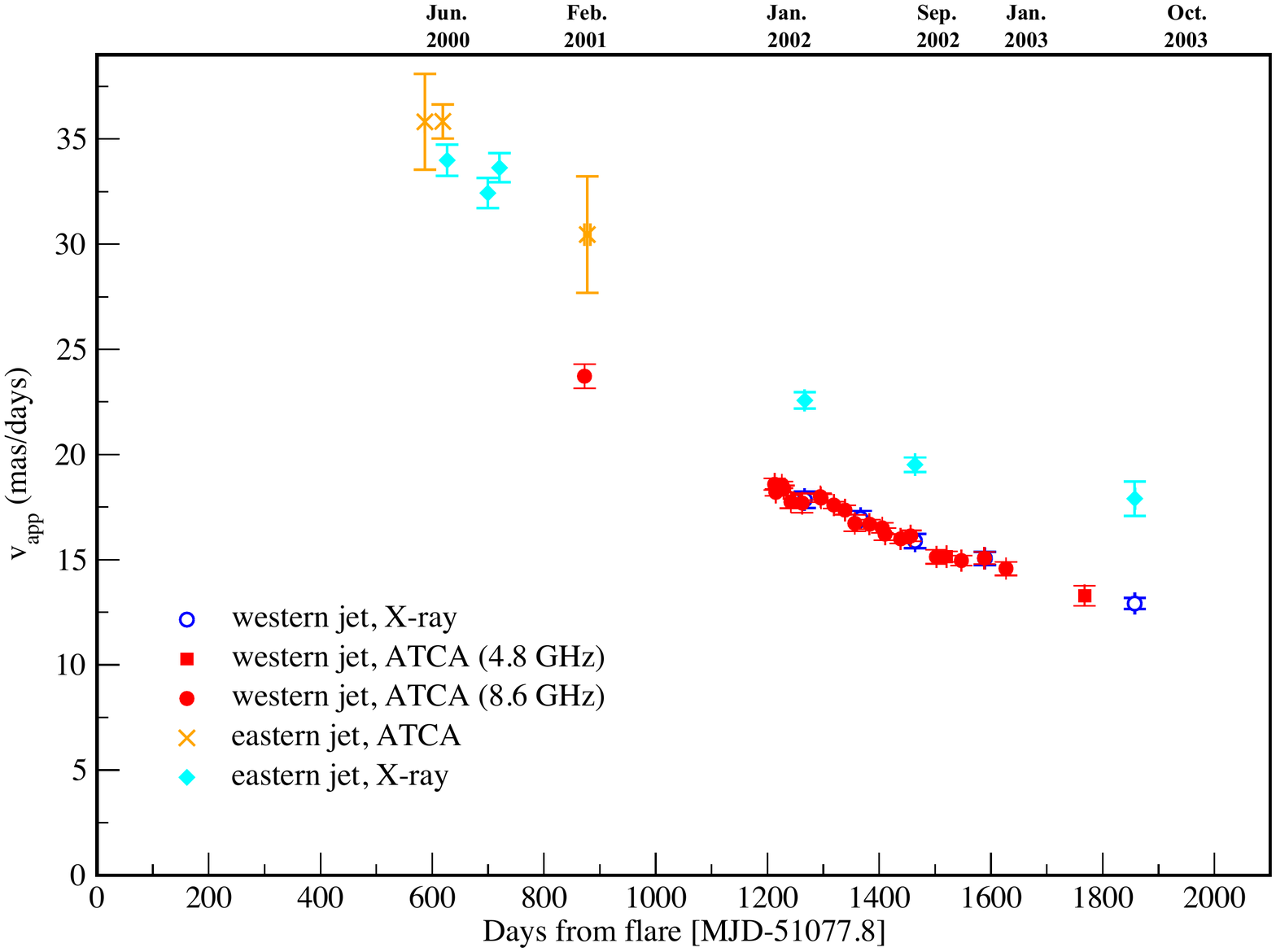}
\caption{Angular separation of the western and eastern jet from XTE J1550$-$564 as a function of time (upper panel) and apparent average velocity with respect to \xte~(lower panel). The X-ray measurements refer to the position and the inferred velocity of the centroid. The horizontal dashed lines in the upper panel mark the limits of the low-density cavity as reported by \citet{SM12}. }
\label{f13}
\end{figure*}

\begin{figure*}
\centering
\includegraphics[scale=0.35]{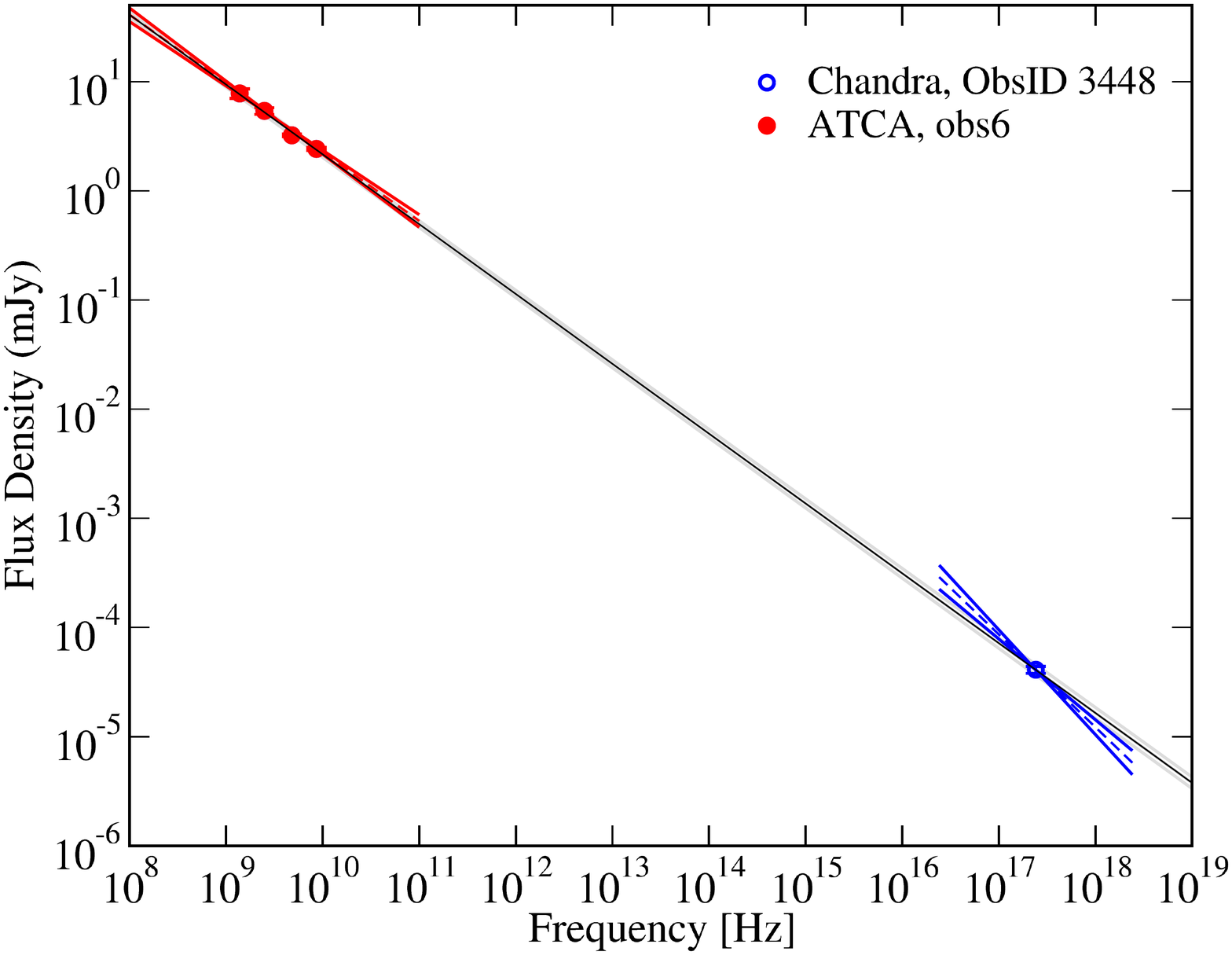}
\includegraphics[scale=0.35]{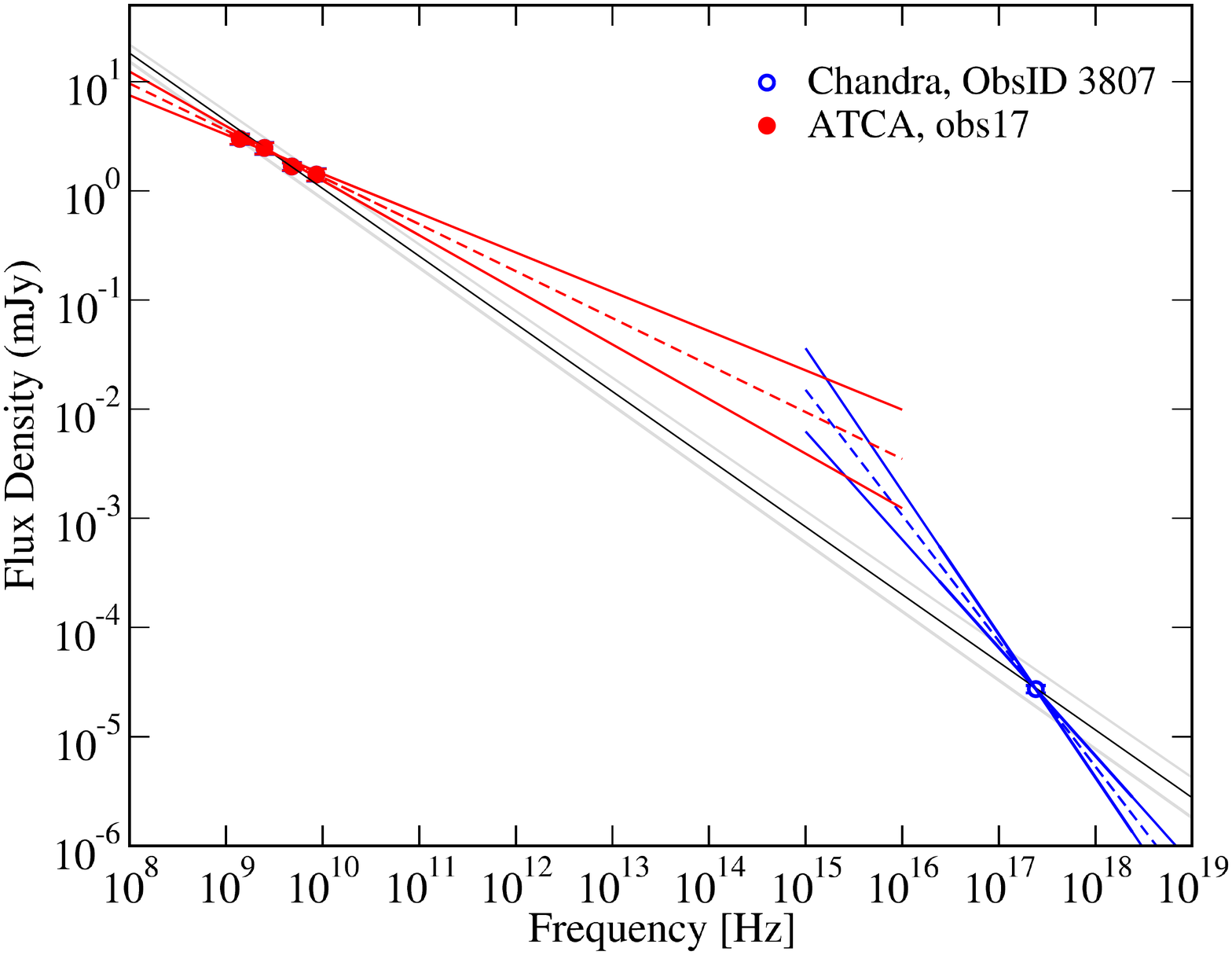}
\includegraphics[scale=0.35]{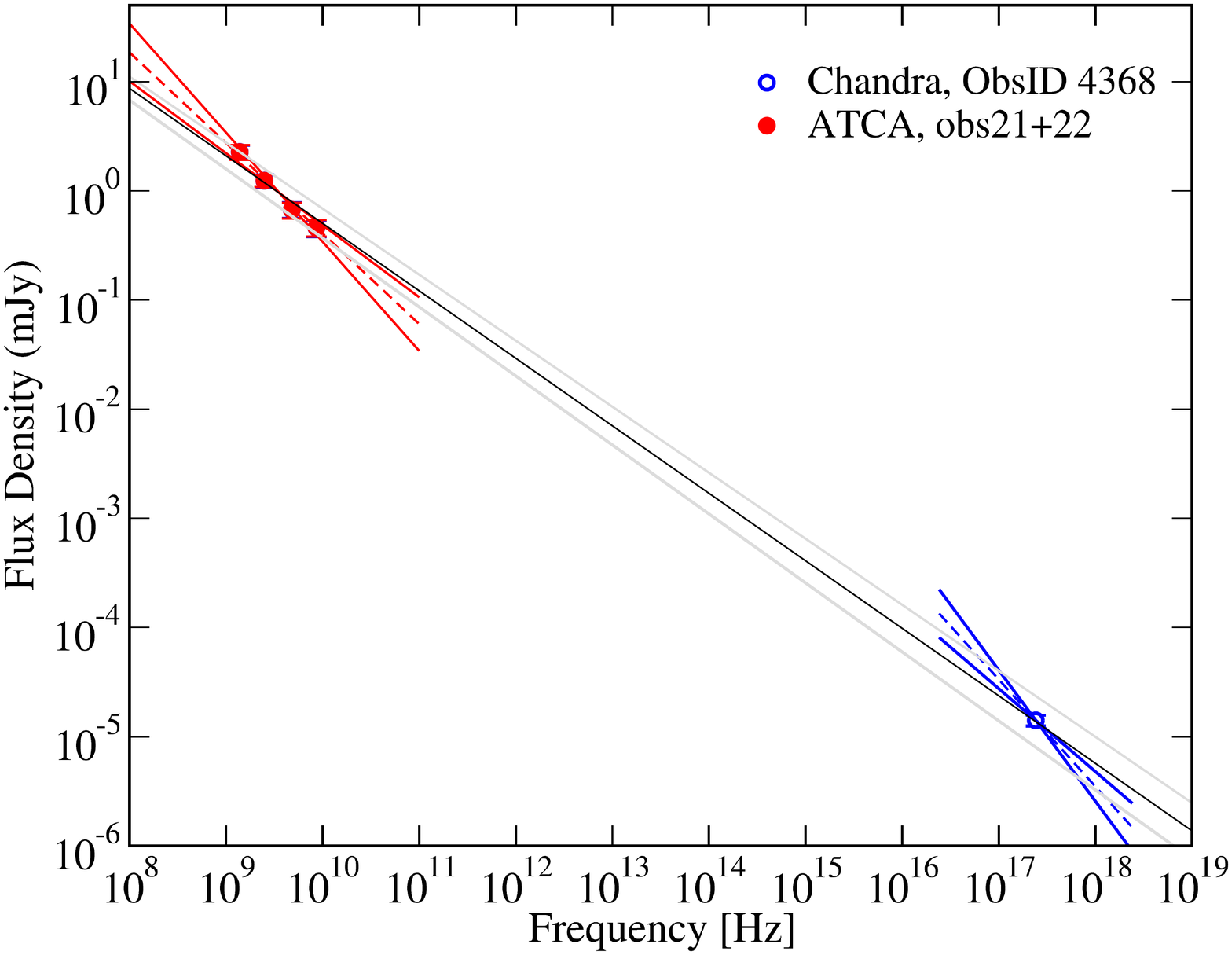}
\caption{Broadband spectrum of the western jet in three epochs: on March 2002 (upper panel), on September 2002 (central panel) and January 2003 (lower panel). The best-fit power-law model (black line) of radio and X-ray data together and uncertainties (grey lines) are also shown.  The fluxes in SEDs always correspond to the total emission of the jet observed at the four radio frequencies and in X-rays at 1 keV.}
\label{f14}
\end{figure*}

\begin{figure*}
\centering
\includegraphics[scale=0.4]{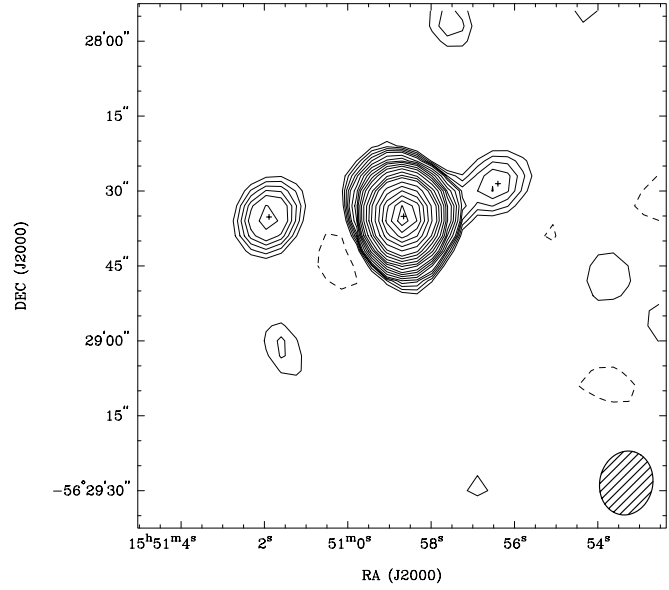}
\caption{ Natural weighted ATCA map at 8.6 GHz of the field of \xte~on February 9 and 20, 2001. Crosses indicate the position of \xte~({\it center}), the eastern ({\it left}) and western ({\it right}) jets. Contours are plotted at $-$3, 3, 4, 5, 6, 7, 9, 11, 13, 15, 18, 21, 25, 30, 35, 40, 50, 60, 70, 80, 90, 100 times the rms noise level of 0.035 mJy beam$^{-1}$. The synthesize beam (lower right corner) is 12.8$\times$10.5 arcsec$^2$.}
\label{f11aa}
\end{figure*}
\clearpage


\appendix
\section{Simulations}

\begin{figure*}
\centering
\includegraphics[scale=0.55]{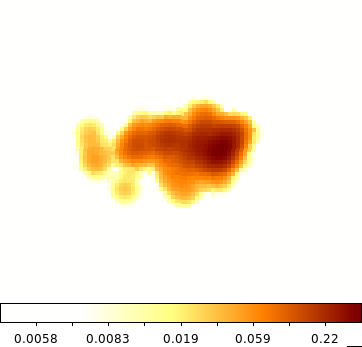}
\includegraphics[scale=0.5]{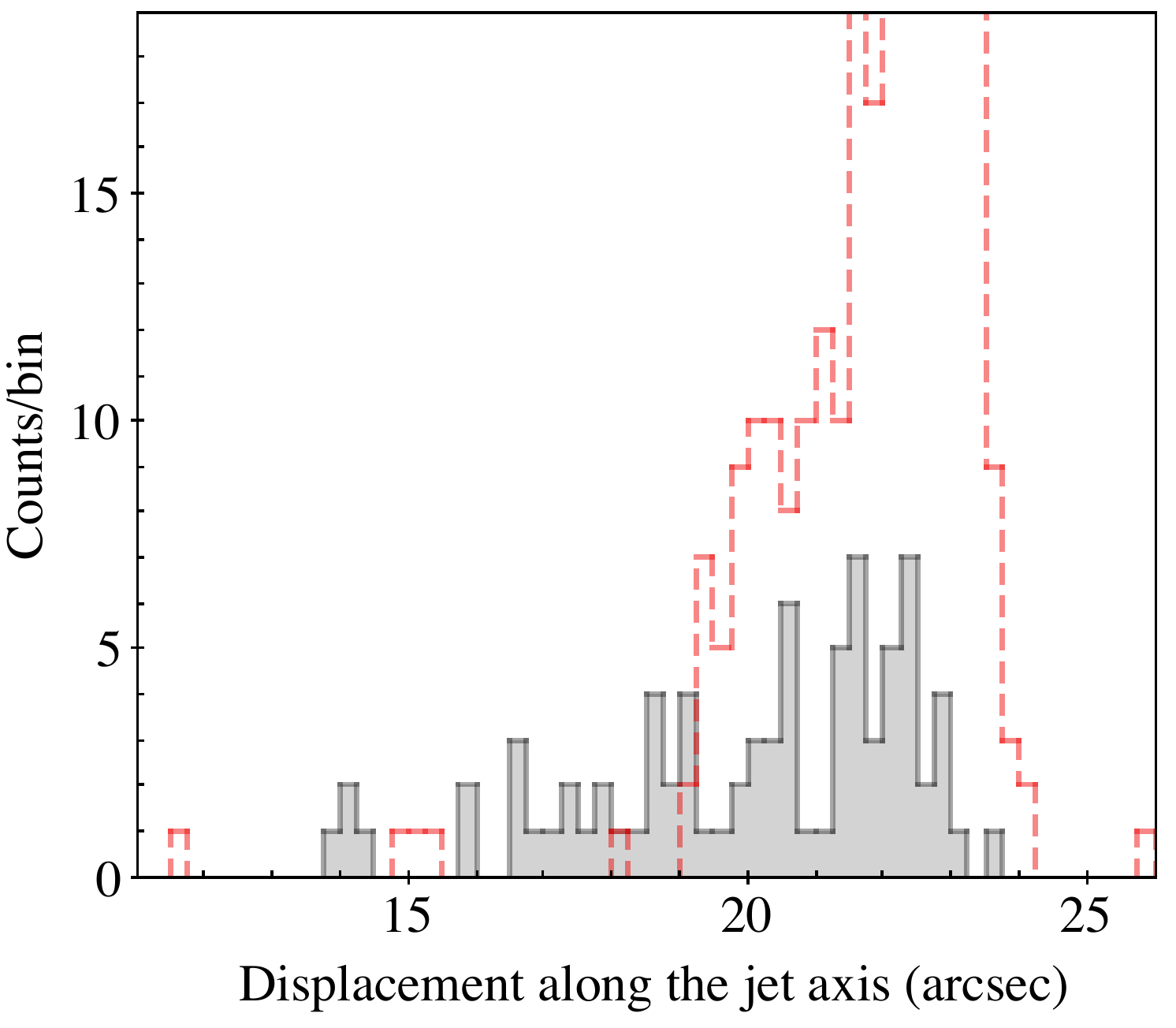}
\caption{ Left panel: example of a MARX simulation of the western X-ray jet.  We used the X-ray image and spectrum of the jet in ObsID 5190 and an exposure time of 24.4 ksec, equal to the observing time in ObsID 3448 (S3 in Table \ref{t6}). The total number of counts in the 0.3-8 keV energy band is 78. The simulated image was smoothed, the pixel size is equal to 0.246\arcsec. Right panel: the longitudinal profile of the simulated jet in the 0.3-8 keV energy range (in grey) is compared with the observed profile of the jet in ObsID 3448. }
\label{f1app}
\end{figure*}

We performed simulations of the X-ray western jet and of the background emission to assess the significance of the expanding X-ray tail.
The first two sets of simulations (S1 and S2 in Table \ref{t6}) aimed at evaluating the uncertainty on the length of the jet measured assuming as the endpoint of the tail: 1- the last bin with $\geq$2 counts attached to the jet body in the longitudinal count profile; 2- the last bin with one count being part of a train of $\geq$5 counts. We considered the first and last \Cha~observations, ObsID 3448 and 5190 respectively, which have the largest difference in angular extension of the counts' longitudinal profiles.  We employed MARX \citep{Dav12} to perform simulations of the jet in each of the two observations. 
The input parameters of the MARX simulations control the spatial distribution and spectral energy distribution of the photons. We used the images of the western jet in each of the two observations for the former and the best fit models of the western jet's spectra (see Table \ref{t3spec}) for the latter.
We used the aspect solution files of the correspondent observations and set the exposure times of the simulations equal to the net (after filtering) exposure times of ObsID 3448 and 5190. 
The X-ray jet was simulated 50 times for each of the two observations (see the left panel of Figure \ref{f1app} for an example). We note that MARX is a Monte Carlo simulation program that operates such that even for the same input parameters two simulations will differ one from the other (unless the parameter controlling the simulation has been opportunely set). This means that the number and spatial distribution of the counts in the 50 simulations vary. We derived the longitudinal count profiles of the simulated jet (see the right panel of Figure \ref{f1app} for an example) and measured the jet extension (with the same criteria adopted for the \Cha~observation). In Table \ref{t6}, we report the mean length of the jet and its standard deviation for the S1 and S2 sets of simulations. The average lengths of the jet of S1 and S2 are 5.6\arcsec\er0.4\arcsec~(6.0\arcsec\er0.4\arcsec~for 1 count/bin) and 8.7\arcsec\er0.4\arcsec~(9.2\arcsec\er0.4\arcsec~for 1 count/bin), respectively. These values are in agreement, within the uncertainties, with the length of the jet observed in ObsID 3448 (5.3\arcsec)  and 5190 (8.3\arcsec/8.8\arcsec). The slight overestimate of the jet length in S2 with respect to the last \Cha~detection is most likely due to the cluster of 5 photons which are $\sim$1.5\arcsec~detached from the jet main body in the longitudinal profiles (see Figure \ref{f4}): while we conservatively decided not to consider them as part of the tail in our measurements, the result of the simulations suggest that they could be spatially related to it.
The result of this first test gives support to our finding of a different length of the jet in the first and last \Cha~observations.\\
\indent
Next, we evaluated the impact on these measurements of the different exposure times of the two observations by simulating the morphology of the jet in ObsID 5190 using the date and aspect solution files of ObsID 3448. Simulation S3 is of a 24.4 ksec observation using the flux and spectrum from ObsID 5190. The average size of the jets is longer in S2 than in S3. 
This is because the number of photons that fall in the terminal region of the tail increases proportionally with the total number of detected photons. 
Nonetheless, the average jet length for S3 is still consistent with the observed jet length in ObsID 5190. Conversely, the average jet size is significantly longer in S3 than in S1 or observed in ObsID 3448. 
We found an average of 10\er4 counts in the tail of the S3 jets beyond the endpoint measured in ObsID 3448 (i.e. the number of counts in the jet at distances larger than 5.8\arcsec~from the leading edge) versus only 2 counts observed in in ObsID 3448. Therefore we would have detected the X-ray tail if it was present in the first Chandra observation, albeit with a slightly reduced length.\\
\noindent
S4 is a simulation of a 24.4 ksec observation using the (brighter) flux and spectrum of ObsID 3448 (and the morphology of ObsID 5190). The average length of the X-ray jet is (9.3\arcsec\er0.4\arcsec/9.6\arcsec\er0.5\arcsec), significantly longer than the size of the jet that we observed in ObsID 3448. Thus, we would have easily detected the tail in the first Chandra observation if it were present and the jet was at the higher flux measured in that observation.

\begin{figure*}
\centering
\includegraphics[scale=0.5]{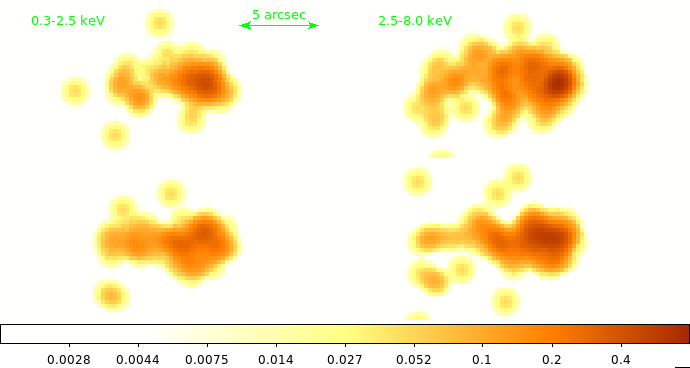}
\caption{ MARX simulations of the western X-ray jet: we used the X-ray image and spectrum of the jet in ObsID 5190 to test the spatial distribution of the emission in the 0.3-2.5 keV and 2.5-8 keV energy bands (see the Appendix for details). Left panel: smoothed 0.3-2.5 keV simulated images of the western jet. We reduced the number of counts in the soft X-ray band (46 cts and 54 cts in the upper and lower panel, respectively) for a comparison with the spatial distribution of the hard X-ray emission. Right panel: smoothed 2.5-8 keV simulated images of the western jet. We increased the number of counts in the hard X-ray band (95 cts and 88 cts  in the upper and lower panel, respectively) for a comparison with emission in the soft X-rays. }
\label{f2app}
\end{figure*}

\indent
We used the bright and compact jet of ObsID 3448 as a ``calibration source'' to understand the uncertainties due to the varying exposure time. We performed simulations of  a 46 ksec long \Cha~observation of the jet of ObsID 3448. The average length of the jet is 5.8\arcsec\er0.3\arcsec~(6.1\arcsec\er0.3\arcsec~for 1 count/bin). An increase of the exposure time to 100 ksec gave consistent results, indicating that the 24 ksec observation was sufficient to define the initial morphology of the jet. 
We did the same test for the jet of ObsID 5190: for an exposure time of 100 ksec we obtained an average length of 8.9\arcsec\er0.4\arcsec, with the minimum and maximum values being 8.3\arcsec~and 9.8\arcsec, respectively. Although the morphology of the extended jet appears more sensitive to the exposure time, the average values of the simulations for 46 and 100 ksec observations are still in agreement.\\ 
\indent
Finally, we investigated the possible impact of the background photons on the measurements of the tail by means of simulations with Sherpa. We selected a rectangular background region in each of the two \Cha images and fitted it with a 2-D constant amplitude model. We used the best fit model to perform 10000 simulations of the background using the \texttt{fake} command in Sherpa. From these simulations, we were able to exclude the possibility of having $\geq$2 background counts per bin and $\geq$5 background counts in 6 consecutive bins (corresponding to a width of 1.5\arcsec, the height of the bin is the same of the bins used for the longitudinal count profiles) at $>99$\% confidence level. 
All in all, the simulations exclude that the (apparent) backward motion of the tail is an artifact due to the different parameters of the \Cha~observations. \\
\indent
The last run of simulations had the goal of evaluating the spatial distribution of the photons in the soft and hard X-ray bands. We used the X-ray image and spectrum of the jet in ObsID 5190, where the emission in the hard X-rays appears more knotty than in the soft X-rays (see Figure \ref{f2}). We tuned the MARX parameters so to obtain a number of soft(/hard) X-ray photons equal to the number of photons detected in the hard(/soft) X-ray band ($\sim$45/90 cts). 
We performed 30 simulations for each energy band. Indeed, the shape of the jet in the hard X-ray band becomes more uniform when the number of counts increases. On the other hand, for a reduced number of photons, the soft X-ray emission appears more knotty (see Figure \ref{f2app}). Thus, based on the available data, we cannot say whether the observed different distribution of photons in the two bands is intrinsic to the jet or an artifact due resulting from having more counts in the soft band. However, we note that common to the two bands, for a comparable number of photons, is the fact that the photons are preferentially accumulated along the central spine.

\begin{table*}
\caption{Western jet: MARX simulations}
\label{t6}
\begin{center}
\begin{tabular}{l c c c c c c}
\hline
\hline
  \multicolumn{1}{c}{Simulations}&
  \multicolumn{1}{c}{Spectrum}&
  \multicolumn{1}{c}{Morphology}&
  \multicolumn{1}{c}{Exposure Time}&
  \multicolumn{1}{c}{Iterations}&
  \multicolumn{1}{c}{Average jet length}&
  \multicolumn{1}{c}{S.D.}\\
\hline
  S1          &3448       &3448    &24.4    &50     &5.6\arcsec/6.0\arcsec      &0.4\arcsec/0.4\arcsec\\    
  S2          &5190       &5190    &46.0    &50     &8.7\arcsec/9.2\arcsec      &0.4\arcsec/0.4\arcsec\\
  S3          &5190       &5190    &24.4    &50     &7.8\arcsec/8.2\arcsec      &0.7\arcsec/0.7\arcsec\\
  S4          &3448       &5190    &24.4    &50     &9.3\arcsec/9.6\arcsec      &0.4\arcsec/0.5\arcsec\\
\hline\end{tabular}
\end{center}
Columns: 1- set of simulations; 2- \Cha~observation from which the input spectrum of the simulation has been extracted; 3- \Cha~observation used for the image file of the simulation; 4- exposure time of the simulation; 5- number of simulations; 6- average angular length of the simulated jets measured from the longitudinal count profiles: the values are for the endpoint set at the 2 count/bin (left) and 1count/bin (right) (see text); 7- standard deviation of the jet average lengths (for the two endpoint criteria).
\end{table*}

%


\bsp	
\label{lastpage}
\end{document}